\documentclass[review]{elsarticle}

\usepackage{lineno,hyperref,amssymb}
\modulolinenumbers[5]    

\journal{Journal of \LaTeX\ Templates}

\usepackage{amssymb}
\usepackage{amsmath}
\usepackage{latexsym}
\usepackage{epsfig}
\usepackage{graphicx}
\usepackage{color}
\usepackage{fouridx}

\makeatletter

\renewcommand*{\p@subsection}{}

\renewcommand*{\p@subsubsection}{}
\makeatother

\newcommand{\clL}{\mathcal{L}}
\newcommand{\clF}{\mathcal{F}}
\newcommand{\half}{\frac{1}{2}}
\newcommand{\clN}{\mathcal{N}}
\newcommand{\clG}{\mathcal{G}}

\newcommand{\clC}{{\cal C}}
\newcommand{\clB}{\mathcal{B}}

\newcommand{\clK}{{\cal K}}
\newcommand{\clT}{\mathcal{T}}
\newcommand{\clD}{\mathcal{D}}
\newcommand{\clU}{\mathcal{U}}
\newcommand{\hh}{\hbar_{\rm ef}}

\newcommand{\hH}{\hat{H}}
\newcommand{\ha}{\hat{a}}
\newcommand{\had}{\hat{a}^{\dag}}

\newcommand{\clM}{{\cal M}}
\newcommand{\clQ}{{\cal Q}}

\newcommand{\prt}{\partial}

\newcommand{\vep}{\varepsilon}

\newcommand{\be}{\begin{equation}}
\newcommand{\ee}{\end{equation}}
\newcommand{\bea}{\begin{eqnarray}}
\newcommand{\eea}{\end{eqnarray}}

\numberwithin{equation}{section}

\begin{document}

\begin{frontmatter}

\title{https://doi.org/10.1142/S0217732321400034 \\
\vline \\
{--------------------------------------------------------------}\\
{\vline}\\
{\vline}\\
FRACTIONAL SCHR\"ODINGER EQUATION \\ IN
 GRAVITATIONAL OPTICS }           

\author{ALEXANDER IOMIN}

\address{ Department of Physics, Technion, Haifa, 32000, Israel \\
iomin@physics.technion.ac.il}

\begin{abstract}
This paper addresses issues surrounding the concept of
fractional quantum mechanics, related to lights propagation
in inhomogeneous nonlinear media, specifically restricted to
a so called gravitational optics. Besides Schr\"odinger Newton equation,
we have also concerned with linear
and nonlinear Airy beam accelerations in flat and curved spaces
and fractal photonics, related to nonlinear Schr\"odinger equation,
where impact of the fractional Laplacian is discussed.
Another important feature of the gravitational optics' implementation
is its geometry with the paraxial approximation, when
quantum mechanics, in particular, fractional quantum mechanics, is
an effective description of optical effects. In this case, fractional-time differentiation reflexes this geometry effect as well.

\end{abstract}

\begin{keyword}
Parabolic equation approximation\sep
Fractional Schr\"odinger equation\sep
Fractional Heisenberg equation\sep
Nonlinear Schr\"odinger equation\sep
Schr\"odinger Newton equation\sep
Airy beam
\end{keyword}

\end{frontmatter}


\section{Introduction}\label{sec:int}

In modern optical experiments of light propagation in
nonlinear composite media a
suitable description can be developed in the framework of
effective quantum equations based on fractional
integro-differentiation. This concept of differentiation of
non-integer orders arises from works by Leibniz,
Liouville, Riemann, Grunwald, and Letnikov, see \textit{e.g.},
Refs. \cite{oldham,Po99}. Its application relates to diffusion - wave
processes with power law distributions. This corresponds to the
absence of characteristic average values for processes
exhibiting many scales
in both space and time \cite{shlesinger,klafter}.
This issue by itself is extremely wide and reflected in vast literature.
In this concise review we discuss only few examples of specific aspects
of fractional quantum mechanics exploring the light propagation
in nonlinear inhomogeneous media,
related to a so called gravitational optics setups \cite{Moti1}.
Gravitational optics effects, considered here in the framework
of fractional calculus, concern primary with linear
and nonlinear Airy beam accelerations in flat and curved spaces
\cite{Moti1,Rivka1} and fractal photonics \cite{Berry79,SeSoDu2012}.
Another important feature of the gravitational optics' implementation
is its geometry with the paraxial approximation, when an effective
quantum mechanical description of the optical effects is valid
\cite{Moti2,LeKrFiSe2012,iom2015}.

As shown in the seminal papers \cite{laskin1,west}, the
appearance of the space fractional derivatives in the
Schr\"odinger equation is natural and relates to the path
integral approach, where fractional concept has been introduced
by means of the Feynman propagator for non-relativistic quantum
mechanics by analogy with fractional Brownian motion
\cite{laskin1,west}. This analogy is natural, since
both are Markov processes \cite{feynman,kac} that
indicates equivalence between the Laplace operators for classical diffusion
equation and the Schr\"odinger equation.
The same, the appearance of the space fractional derivatives in
the Schr\"odinger equation is natural, since both the standard
Schr\"odinger equation and the space fractional one obey the
Markov chain rule.
Technically, it is described by a fractional Laplacian
in the form of the Riesz fractional derivative
(see \ref{sec:app_A})
\begin{equation}\label{Riesz}
(-\Delta)^{\frac{\alpha}{2}}\equiv
\fourIdx{}{\infty}{\alpha}{\lvert x\rvert}{\mathcal{D}}f(x)
=\frac{1}{2\cos\tfrac{\alpha\pi}{2}}
\left[\fourIdx{}{-\infty}{\alpha}{x}{\mathcal{D}}f(x)
+\fourIdx{}{-\infty}{\nu}{-x}{\mathcal{D}}f(-x)\right]\, ,
\end{equation}
where $\alpha\in (0,2]$.
This introduction of L\'evy flights due to the power law kernel,
as well as the L\'evy measure in quantum mechanics is
based on the generalization of the self-consistency condition,
known as the Bachelor-Smoluchowski-Kol\-mo\-go\-rov chain equation
(or the Einstein-Smoluchowski-Kol\-mo\-go\-rov-Chapman equation,
see \textit{e.g.}, \cite{shlesinger}), established for the Wiener
process for the conditional probability $W(x,t|x_0,t_0)$
\be\label{Smoluch} %
W(x,t|x_0,t_0)=\int_{-\infty}^{\infty}W(x,t|x',t')W(x',t'|x_0,t_0)\,
dx'\,  . \ee%
In the case of the translational symmetry, it reads
$W(x,t|x_0,t_0)=W(x-x_0,t|t_0)$. Straightforward generalization of
this expression by the L\'evy process is expressed through the
Fourier transform
\be\label{levy1}  %
W(x,t|x_0,t_0)=\int_{-\infty}^{\infty}
e^{ik(x-x_0)}e^{-K_{\alpha}t|k|^{\alpha}}\,dk\, ,  \ee %
where $0<\alpha\leq 2$ and $K_{\alpha}$ is a generalized diffusion
coefficient \cite{klafter}.
Eventually, Eq. \eqref{levy1} in the form of a generalized Feynman-Kac
formula \cite{laskin1,ZT} results from the fractional space
Schr\"odinger equation (FSSE), which in dimensionless form reads
\begin{equation}\label{fse-1}
i\hh\partial_t\psi(x,t)=(\hh^{\alpha}/2)(-\Delta)^{\frac{\alpha}{2}}\psi(x,t)
+V(x)\psi(x,t)\, ,
\end{equation}
where $V(x)$ is a potential field, while $K_{\alpha}$ is replaced by
$(-i\hh)^{\alpha-1}/2$ with $\hh$ being an effective dimensionless
Planck constant.

Recently, nonlocal, fractional mechanics has been demonstrated
experimentally \cite{Longhi}. This physical
implementation of space-fractional Schr\"odinger equation is based
on transverse light dynamics in
aspherical optical resonators that realizes the fractional quantum
harmonic oscillator \cite{Laskin2} (also known as massless relativistic
quantum oscillator \cite{relat_oscill})
in which dual Airy beams can be generated. Further ``fractionalization''
of optical beams attracts much attention in both theoretical
and numerical studies of both linear \cite{GVega} and nonlinear
\cite{Y_Zhang,L_Zhang2,L_Zhang1}
Schr\"odinger equations. Another important achievement relates to the
fabrication and exploration of optical \cite{HuMcD2005,Sha2002}
and quantum  fractals \cite{jad2014,SiSu2019}.
Then the fractional Laplacian \eqref{Riesz} plays important role in both
quantum and wave-diffusion processes \cite{SiSu2019}
(see Sec~\ref{sec:slab}).

Fractional time Schr\"odinger equations
(FTSE) are another large class of quantum problems
with long memory effects.
This kind of non-Markovian, non-unitary quantum dynamics has been
observed in real physical systems \cite{wo2010,iom2009,IS16}.
It also corresponds to non-Hermitian
extension of quantum mechanics \cite{Bender}
and attracts a growing interest in optics, see, \textit{e.g.},
\cite{OPTICS} (and references therein).
However, an introduction of this fractional time concept in
quantum mechanics is not an easy task and needs a
special care. This situation is reflected in recent reviews
\cite{Ta08,La2017,iom2019}. A FTSE has been introduced by
analogy with a fractional Fokker-Planck equation (FFPE)
by means of the analytic continuation of time
$t\rightarrow -it/\hbar$ \cite{naber}, where $\hbar$ is the Planck constant.
This approach attracts much attention and has been extensively studied,
see \textit{e.g.}, Refs. \cite{list1,list2,iom2011,list3,list4}
(and references therein).
Another analytic continuation of time has been
suggested as well
\cite{IS16,AcYaHa13,SaPeLe14} by
$t\rightarrow t/(i\hbar)^{1/\beta}$ with $0<\beta<1$.
At these definitions\footnote{In each case the Planck constant
$\hbar$ has a special dimensionality definition. However, in this
presentation we use an effective dimensionless Planck constant $\hh$,
introduced above. Note also that this procedure for the
dimensionless variables can be always performed,
see Refs.~\cite{naber,AcYaHa13}.},
a standard partial derivative of the wave function
with respect to time is replaced by a so called Caputo
fractional derivative,
which is a convolution integral of the wave function with a
power law kernel:
\begin{equation}\label{int-1}
\partial_t\psi(t)\rightarrow\partial^{\beta}_t\psi(t)\equiv
\clD_{C}^{\beta}\psi(t)=
\frac{1}{\Gamma(1-\beta)}\int_0^t(t-\tau)^{-\beta}
\frac{d}{d\tau}\psi(\tau)d\tau\, ,
\end{equation}
where $\Gamma(\beta+1)=\beta\Gamma(\beta)$ is the gamma function,
see \ref{sec:app_A}.
Contrary to the space fractional concept of Eq. \eqref{fse-1},
the fractional time
quantum mechanics is non-unitary. That is, it
violates the Stone's theorem on the
one-parameter unitary group\footnote{Stone's theorem establishes a group
property $\hat{U}(t)\hat{U}(s)=\hat{U}(t+s)$
 for the evolution (unitary) operator $\hat{U}(t)$.} \cite{stone}
and correspondingly the Feynman path integral approach
does not exist.

A general formulation of a nonlinear fractional
Schr\"odinger equation (nonlinear FSE) that envisages possible experimental
realizations in optics can be formulated as the generalization
of Eqs. \eqref{fse-1} and \eqref{int-1} as follows
\begin{multline}\label{gfse-2}
i\hh\partial_t^{\beta}\psi(x,t)
=(\hh^{\alpha}/2)(-\Delta)^{\frac{\alpha}{2}}\psi(x,t) +
V(x,t)\psi(x,t) + \\
+\clB\psi(x,t)\int \clQ(x-x')|\psi(x',t)|^2dx'\, .
\end{multline}
Here $\beta\in (0,1]$, $\alpha\in (0,2]$, while
$\clB$ is a nonlinearity parameter
and $\clQ(x-x')$ is a power law kernel.
An effective potential $V(x,t)$ results from the optical setup.
When $\clQ(x-x')=\delta(x-x')$,
Eq. \eqref{gfse-2} is a fractional generalization of the nonlinear
Schr\"odinger equation (NLSE). The latter is a well known equation
with a variety of applications, including nonlinear optics, see
recent review \cite{Fibich}. In optics, $t\equiv z\in [0,\infty)$
in Eq. \eqref{gfse-2} is usually
an effective time, which results from the paraxial
description of the Helmholtz equation
and corresponds to the longitudinal coordinates $z$
\cite{Rivka1,Moti2}, while $x\in\mathbb{R}^n$ is position coordinates in the $n$-dimensional space.

\section{FSE in slab geometry: parabolic equation approximation}
\label{sec:slab}

As admitted above, the Schr\"{o}dinger equation in optics appears
as a formal effective description of diffusive wave transport in
complex inhomogeneous media in the parabolic equation approximation
that corresponds to the paraxial small angle approximation.
Therefore, one can obtain the FSE as an effective way
to solve fractional eigenvalue problem in slab geometry, where the paraxial
small angle approximation is naturally applied.
The method of parabolic equation approximation was
first applied by Leontovich in study of radio-waves spreading
\cite{leontovich}
and later developed in detail by Khokhlov \cite{khohlov} (see also
Refs.~\cite{kliatskin,tappert}).

In the section, our main concern is the Helmholtz fractional
equation, which relates to the L\'evy process in
the 2D slab geometry described by
dimensionless $(x,z)$ variables, where $z\in (0,\,\infty)$ and
$x\in [-L,\,L]$. Therefore the wave function $\Psi(z,x,\omega)$
is determined by the fractional Helmholtz equation \cite{IS16}
\begin{equation}\label{ffpe_3}
\partial_z^{\alpha}\Psi+\mathcal{D}_{|r|}^{\alpha}\Psi+
\omega\Psi=0\, ,
\end{equation}
where $\omega$ is the propagating wave/heat frequency.
It should be stressed that fractional space derivatives in Eq.
\eqref{ffpe_3} describe L\'evy flights \cite{klafter}.
In particular, we specify here optical ray dynamics
in L\'evy glasses \cite{LevyLens}, where 
L\'evy flights\footnote{L\'evy
glasses are specially prepared optical
material in which the L\'evy flights are controlled by the power law
distribution of the step-length of a free ray dynamics, which can be
specially chosen in the power law form $\sim 1/l^{\alpha+1}$.}
can be described by Eq. \eqref{ffpe_3}.
Another interesting phenomenon, which is described by Eq. \eqref{ffpe_3},
is superdiffusion of ultra-cold atoms in optical lattices
\cite{nirD}\footnote{There are L\'evy walks, and the theoretical
explanation of this fact, presented within the standard semiclassical
treatment of Sisyphus cooling \cite{zoller,eli1}, is based on a
study of the microscopic characteristics of the atomic motion in optical
lattices and recoil distributions resulting in macroscopic L\'evy walks in
space, such that the L\'evy distribution
of the flights depends on the lattice potential depth \cite{zoller}.
The flight times and velocities of atoms are coupled, and these relations,
established in asymptotically logarithmic potential, have been studied for
different regimes of the atomic dynamics \cite{eli1,eli2},
so the cold atom problem is a variant of the L\'evy walks,
see also discussion of the L\'evy flights
in the framework of a superdiffusive comb model \cite{iom2012}.}

We use the Caputo fractional derivative
with respect to the longitudinal coordinate $z\in [0,\,\infty)$,
namely
$\partial_z^{\alpha}\equiv \fourIdx{C}{0}{\alpha}{z}{\mathcal{D}}$,
while the Riesz fractional derivative
$\mathcal{D}_{|x|}^{\alpha}\equiv
\fourIdx{}{L}{\alpha}{\lvert x\rvert}{\mathcal{D}}=
\fourIdx{RL}{-L}{\alpha}{x}{\mathcal{D}}+
\fourIdx{RL}{x}{\alpha}{L}{\mathcal{D}}$ is used
for the orthogonal direction $x\in[-L,\, L]$.
When the size $L$ is less than the L\'evy flight lengths in the
longitudinal direction, the transport is of a small grazing
angle with respect to the longitudinal direction.
The solution $\Psi(z,x,\omega)$ can be presented in the
following multiplication form
\begin{equation}\label{ffpe_4}
\Psi(z,x,\omega)=e^{ikz}\psi(z,x,\omega).
\end{equation}
Substitution Eq. (\ref{ffpe_3}) in Eq. \eqref{ffpe_4} yields
the following integration
\begin{equation}\label{ffpe_5}
\partial_{z}^{\alpha}\Psi(z,x,\omega)=\frac{1}{\Gamma(2-\alpha)}
\int_0^z(z-z')^{2-\alpha-1}\frac{d^2}{dz'^2}\left[e^{ikz'}
\psi(z',x,\omega)\right]dz'\, .
\end{equation}
Note that the ``initial'' conditions at $z=0$ for both $\Psi$
and $\psi$ are the same: $\psi(z=0)=\Psi(z=0)=\Psi_0(x,\omega)$.

Now the parabolic equation in the paraxial approximation can be obtained.
Taking into account that $\psi(z,x,\omega)$ is a slowly-varying
function of $z$, such that
\begin{equation}\label{ffpe_5_6}
\Big|\frac{\partial^2 \psi}{\partial z^2}\Big|\ll
\Big|2k \frac{\partial \psi}{\partial z}\Big|\, ,
\end{equation}
one obtains
\begin{equation}\label{ffpe_6}
 \frac{d^2}{dz^2}\left[e^{ikz}\psi(z,x,\omega)\right]\approx
 2ike^{ikz}\frac{d}{dz}\psi(z,x,\omega)\, .
 \end{equation}
After substitution of this approximation in Eq. (\ref{ffpe_5}), one obtains
from Eqs.~(\ref{ffpe_3}), (\ref{ffpe_4}), and (\ref{ffpe_5})
\begin{equation}\label{ffpe_7}
2ik \fourIdx{}{0}{2-\alpha}{z}{I}
\left[e^{ikz}\partial_z\psi(z,x,\omega)\right]+
\mathcal{D}_{|x|}^{\alpha}\psi(z,x,\omega)e^{ikz}+
\omega\psi(z,x,\omega)e^{ikz}=0\, .
\end{equation}
To remove the exponential $e^{ikz}$ from Eq.~(\ref{ffpe_7}),
we insert it inside the derivative $e^{ikz}\partial_z\psi(z,x,\omega)=
\partial_z[e^{ikz}\psi(z,x,\omega)]+
{\rm o}(k)$. The term ${\rm o}(k)=ik e^{ikz}\psi(z,x,\omega)$
can be neglected in Eq.~(\ref{ffpe_7}) since it is of
the order of ${\rm O}(k^2)$, which must be
neglected for $k\ll 1$. Note also that
$\fourIdx{}{0}{2-\alpha}{z}{I}\frac{d}{dz}f(z)=
\partial_z^{\beta}f(z)$, where
$\beta=\alpha-1$ and $0<\beta<1$. In general case,
the order of the Caputo
fractional derivative in Eq. \eqref{ffpe_3} can be $\alpha'\neq \alpha$
with $(0<\alpha'\leq 2)$, then $\beta=\alpha'-1$ does not relate to
$\alpha$. The Laplace transform can be performed:
$\mathcal{L}[e^{ikz}\psi(z)]=\tilde{\psi}(s-ik)$.
Therefore, one obtains from Eq.~(\ref{ffpe_7})
\begin{equation}\label{ffpe_8}
2ik[s^{\beta}\tilde{\psi}(s-ik)-s^{\beta-1}\psi(z=0)]+
\mathcal{D}_{|r|}^{\alpha}\tilde{u}(s-ik)+
\omega_{\beta}\tilde{u}(s-ik)\, .
\end{equation}
Performing the shift $s-ik\rightarrow s$ and neglecting again the terms of
the order of ${\rm o}(k)$ in Eq.~(\ref{ffpe_8}), and then performing the
Laplace inversion, one obtains the Helmholtz equation in
the form of the effective fractional Schr\"odinger equation (FSE),
where the $z$ coordinate plays the role of an effective time
\begin{equation}\label{ffpe_9}
2ik\partial_z^{\beta}\psi+\mathcal{D}_{|r|}^{\alpha}\psi+\omega\psi=0\, .
\end{equation}
The ``initial'' condition at $z=0$ corresponds to the boundary condition for
the initial problem in Eq.~(\ref{ffpe_3}).
If one supposes that there is a source of the signal at $z=0$,
then the initial condition is $\psi(z=0,x)=\Psi_0(x,0)$.
The boundary conditions at $x=\pm L$ are $\psi(x=\pm L,z)=0$.

\section{Fractional space Schr\"odinger equation}\label{sec:part1}

Taking $\beta=1$ in the generalized FSE
\eqref{gfse-2}, we consider various realizations
of the light propagation in nonlinear composite materials describing
the phenomenon in the framework of the fractional NLSE,
including Airy beams with a metric determinant different from unite.
This so-called gravitational optics can be also realized in the form of
a Schr\"odinger - Newton equation. Another  realization
of Eq. \eqref{gfse-2}
corresponds to the NLSE in random optical potential.
An extension method, which is relevant to the quantum
field theory consideration is discussed as well.

\subsection{Airy-Fox beams}
\label{sec:foxba}

Experimental advances in Airy beam acceleration
has been demonstrated in curved optical spaces
in both linear and nonlinear optics
with predesigned refractive index varying so as to
create curved space environment for light, see an extended
discussion in Ref.~\cite{Rivka1}.
Therefore, describing this process in the framework of the
generalized FSE \eqref{gfse-2}, the latter
according to Refs. \cite{Rivka1,L_Zhang1} becomes the
fractional NLSE as follows
\begin{equation}\label{beam-1}
i\hh\partial_t\psi(x,t)
=\frac{(\hh)^{\alpha}}{2g(t)}(-\Delta)^{\frac{\alpha}{2}}\psi(x,t) -
\frac{\clB}{g(t)}|\psi(x,t)|^2\psi(x,t)\, .
\end{equation}
Here $g=g(t)$ is a metric determinant (determinant of a metric tensor)
that reflects a curved optical space. Potential $V(t)$ in Eq. \eqref{gfse-2}
is a function of time and can be omitted due a gauge transformation
$\psi(x,t) \leftrightarrow e^{-(i/\hh)\int_0^t V(\tau)d\tau}\psi(x,t)$.
The nonlinear term is due to the Kerr effect and the effective
nonlinear coefficient now is $\clB\rightarrow -\clB/g(t)$
(see Eq. (16) in Ref. \cite{Rivka1}).
This nonlinear equation stands for numerical investigations
\cite{L_Zhang2,Rivka1}.
For $\alpha=2$, the solution of the linear part of Eq. \eqref{beam-1}
corresponds to the solution in the form of an accelerating Airy beam,
which propagates along the curve $x=g_1^2(t)$ at the condition
$|\psi(x,0)|=|\psi(x-g_1^2,t)|$.
This curve is determined by the metric tensor according to
$g_1=g_1(t)=\frac{1}{\hh}\int_0^tg^{-1}(t')dt'$
\cite{Rivka1}. This longstanding problem attracts much attentions
since the seminal result \cite{BeBa79}. In particular,
the linear Airy beam solution \cite{Rivka1}
(for $\alpha=2$) in the form of the Fourier transformation
reads
\begin{multline}\label{beam-4}
\hat{\psi}_{\rm Ai}(k,t)=\int_{-\infty}^{\infty}
e^{-ikx} \frac{1}{\pi}\int_0^{\infty}d\kappa
e^{i\kappa^3/3+i\kappa X(x,t)}e^{ig_1(t)X(x,t)} \\
=2\exp\left\{i[k-g_1^2(t)]^3/3-ikg_1^2(t)\right\} \, ,
\end{multline}
where $X(x,t)\equiv x-g_1^2(t)$.

The linear part of the fractional NLSE \eqref{beam-1}
\begin{equation}\label{beam-1a}
i\hh\partial_t\psi(x,t)
=\frac{(\hh)^{\alpha}}{2g(t)}(-\Delta)^{\frac{\alpha}{2}}\psi(x,t)
\end{equation}
does not correspond to the Airy beam.
For $\alpha\neq 2$ the Green's function is expressed in the form
of the Fox $H$-function as follows
\begin{multline}\label{beam-2}
\clG(x,t)=\frac{2}{\pi}\int_0^{\infty}\cos(kx)
\exp\left[-\frac{i}{2}\hh^{\alpha-1}g_1(t)|k|^{\alpha}\right] \\
=\frac{2}{\pi}\int_0^{\infty}\cos(kx)
H_{0,1}^{1,0}\left[i\frac{\hh^{\alpha-1}}{2}|k|^{\alpha}g_1(t)
\left|\begin{array}{cc} - \\ (0,1)\end{array}\right.\right]dk \\
=\frac{2}{x}
H_{2,2}^{1,1}\left[-i\frac{2}{g_1(t)\hh^{\alpha-1}}|x|^{\alpha}
\left|\begin{array}{cc} (1,1),(1,\frac{\alpha}{2})\\
(1,\alpha),(1,0),(1,\frac{\alpha}{2})\end{array}\right.\right] \, ,
\end{multline}
where $\hat{\psi}_{\rm Fox}(k,t)=
2\exp[-(i\hh^{\alpha-1}/2)g_1(t)|k|^{\alpha}]$
is the Fourier image of the Fox beam \eqref{beam-2}, see \ref{sec:app_A}.

It is instructive to consider the dynamics \eqref{beam-1a}
for the initial Airy function.
\begin{equation}\label{beam-6}
\psi(x,0)=Ai(ax/\hh^{2/3})\, ,
\end{equation}
where $a>0$ is arbitrary. For simplicity sake, we take
$a=\hh^{2/3}$. Therefore,
we have that the solution to \eqref{beam-1a} is \cite{BeBa79}
\begin{equation}\label{beam-7}
\psi_{\rm lin}(x,t)= \int
e^{ikx}e^{-(i\hh^{\alpha-1}/2)g_1(t)|k|^{\alpha}+ik^3/3}dk
=\int \clG(x-x',t)Ai(x')dx'\, .
\end{equation}

Let us consider the asymptotic solution for $t\rightarrow\infty$.
If along the time, $g_1(t)$ increases, then
the argument of the Fox $H$-function
in Eq. \eqref{beam-2} tends to zero. The Fox $H$-function
satisfies the asymptotic theorem for small arguments \cite{MSH2010}
that yields $H_{m,n}^{p,q}(z)={\rm O}(z^c)$ for $|z|\rightarrow 0$.
In our case, $c=1/\alpha$. Therefore, the wave function is independent of
$x$ and is a function of time only in the large time asymptotics.
In particular, in the flat space, when $g=1$, the wave function is
\begin{equation}\label{beam-3}
\psi_{\rm lin}(x,t)\sim 2(1-i)[g_1(t)]^{-\frac{1}{\alpha}}=
2(1-i)t^{-\frac{1}{\alpha}}\, .
\end{equation}
Then the nonlinear term in Eq. \eqref{beam-1}
is small enough for a
perturbation theory consideration.

\subsection{L\'evy flights in NLSE}\label{sec:fnlse}

Disregarding the effective potential $V(x,t)$ in Eq. \eqref{gfse-2}
and considering the kernel $\clQ=\delta(x-x')$, we arrive at the
fractional NLSE
\begin{equation}\label{fnlse-1}
i\hh\partial_t\psi(x,t)
=(\hh^{\alpha}/2)(-\Delta)^{\frac{\alpha}{2}}\psi(x,t)
+\clB|\psi(x,t)|^2\psi(x,t)\, .
\end{equation}
Fractional NLSE attracts much attention in both physical
and mathematical literature. Among the vast literature
we only mention that it has been appeared in the form
of the continuous limit of the dynamics of nonlinear lattices with
long range interactions \cite{LaZa2006,Ta2015} and
it also stems from the fractional generalization of the
Ginzburg-Landau model
\cite{TaZa2005,MiRa2005,TaZa2006a,TaZa2006b,ZaEdTa2007}.
The latter case corresponds to the fractional analogy with
the free energy expansion
\begin{equation}\label{free-energy}
F=F_0+\int  \left[|\nabla^{\alpha}\psi(x)|^2 +|\psi|^2+
\half|\psi|^4\right]dx\,
\end{equation}
that results from the power low of the order parameter  $x^{-\alpha}$	
for the coexisting nonlocal symmetry \cite{MiRa2005}.

\subsubsection{Extension method}\label{sec:em}
Another elegant way introducing
the fractional Laplacian has been suggested in mathematical
literature \cite{CaSi2007} as a fractional generalization of an
extension method for the square root of the Laplacian,
$(-\Delta)^{\half}f({\bf x})$, or half Laplacian\footnote{In
Ref.~\cite{CaSi2007}, the fractional Laplace operator
is defined in the following regularized form
$(-\Delta)^{\alpha/2}(f(x)=\frac{1}{C_{n,\alpha}}
\int\frac{f(x)-f(\xi)}{|x-\xi|^{n+\alpha}}
d\xi=\frac{1}{C_{n-1,\alpha}}\int\frac{f'(\xi)d\xi}
{|x-\xi|^{n-1+\alpha}}$.}.
As admitted in Ref. \cite{PaKoetal2020}, it is the ``revolutianary''
result which demonstrates that the fractional Laplacian can be
expressed via the Dirichlet to Neumann map
associated with a particular extension problem. Namely,
it has been shown that the fractional
Laplacian $(-\Delta)^{\frac{\alpha}{2}}f({\bf x})$
can be related to solutions
of an extension problem as follows \cite{CaSi2007}. For a function
$f : ~ \mathbb{R}^n \rightarrow \mathbb{R}$,
there is the extension
$u :~ \mathbb{R}^n\times [0,\infty) \rightarrow\mathbb{R}$
that satisfies the equation
\begin{equation}\label{fnlse-2}
\Delta_xu + a y^{-1}\partial_yu + \partial_y^2u = 0\, ,
\end{equation}
with the Dirichlet boundary condition $u(x,0)=f(x)$.
Here $a=1-\alpha$ and $x\equiv{\bf x}$ is a $n$ dimensional vector
in $\mathbb{R}^n$.
Then the fractional Laplacian is defined as the Neumann boundary
condition to the extension problem \eqref{fnlse-2}, namely it reads
\begin{equation}\label{fnlse-2a}
C_{n,\alpha}(-\Delta_x)^{\frac{\alpha}{2}}f=-\lim_{y\to 0}y^a\partial_yu
=\frac{1}{\alpha}\lim_{y\to 0}\frac{u(x,y)-u(x,0)}{y^{\alpha}}\, ,
\end{equation}
where $C_{n,\alpha}$ is a constant multiplier.

Let us consider half Laplacian.
When $a=0$, the stationary NLSE
with the half Laplacian reads \cite{CaTa2010,Ta2011}
$(-\Delta)^{1/2}u=w(u)+eu$ with $u(x)>0$ in the domain
$\Omega\in \mathbb{R}^m$ and
$u=0$ at the boundary  $\partial\Omega$, and $w(u)\equiv u^3$,
in our case. The extension problem
in a half cylinder $\Omega\times[0, \infty)$ is
$\Delta v=0$ (or $-\Delta_x v=\partial_y^2v$) with $v(x,y)>0$,
and the boundary conditions now are
$v=0$ at $\partial\Omega\times[0, \infty)$ and
$\partial_{\bf l}v(x,y)|_{y=0}=w(v)|_{y=0}$. Here ${\bf l}$
is the unit outer normal to $\Omega\times{0}$.

The fractional, half Laplacian is determined by the extension problem.
Applying twice the operator
$\clT : ~u \rightarrow -\partial_yv(x,0)$,
one obtains for the harmonic function and its derivative:
$\clT \circ\clT u = \partial_y^2 v |_{y=0}= -\Delta_xv |_{y=0}
= -\Delta u$. Therefore $\clT=(-\Delta)^{\half}$
is the half Laplacian and in the same time it is
the Neumann boundary condition $-\partial_yv(x, 0)$
for the extension problem.

\subsubsection{Fractional Laplacian and wave equation in curved space}

Let us discuss Eq. \eqref{fnlse-2} with respect to
the experimental setup
in curved optical spaces of Sec.~\ref{sec:foxba}.
To this end we simplify the consideration for the
electric field function $\psi(x,t)$ in Eq. \eqref{beam-1a}
and consider it as a harmonic function,
which satisfies the wave equation \eqref{fnlse-2} as follows
\begin{equation}\label{fnlse-3}
\partial_t^2\psi(x,t)+\frac{1}{g(t)}\Delta_x\psi(x,t)=0\, .
\end{equation}
The initial condition, as the Dirichlet boundary condition,
is $\psi(x,0)= f(x)$.
The metric determinant is chosen in the form
$g(t)=t^{\frac{1-\alpha}{\alpha}}$.
Zero boundary conditions are taken at $x=\pm\infty$.
Performing the change of the variables $t=(\tau/\alpha)^{\alpha}$ and
$\partial_t\psi=\tau^{1-\alpha}\partial_{\tau}\psi$,
one obtains the Euler-Lagrange equation \eqref{fnlse-2}
\begin{equation}\label{fnlse-4}
\Delta_x\psi+\frac{1-\alpha}{\tau}\partial_{\tau}\psi+
\partial_{\tau}^2\psi=0\, ,
\end{equation}
which can be produced from the functional (Lagrangian)
$\int |\nabla\psi|^2\tau^{1-\alpha}dxd\tau$,
as well \cite{CaSi2007}.

As obtained in Ref.~\cite{CaSi2007}, the solution
to Eq. \eqref{fnlse-3} (or Eq. \eqref{fnlse-2})
has the power law form when $n-1+a>1$, which is
determined by a Poisson formula
\begin{equation}\label{fnlse-5}
\psi(x,\tau)=\int_{\mathbb{R}^n} \clG(x-x',\tau)f(x')dx'\, .
\end{equation}
The Poisson kernel, or the Green's function reads
\begin{equation}\label{fnlse-6}
\clG(x,\tau) = \frac{C_{n,\alpha}\tau}
{\left[|x|^2 + \alpha^2 \tau^{2/\alpha}\right]^{(n+\alpha)}}\, ,
\end{equation}
where the constant $C_{n,\alpha}=C_{n+\alpha}$ is
$C_{k} = \pi^{k/2}\Gamma(k/2-1)/4$.
The Poisson formula \eqref{fnlse-5} establishes also the relation
between the solution $\psi(x,\tau)$ and the fractional Laplacian
in the form of the Neumann boundary condition
\eqref{fnlse-2a}, which reads \cite{CaSi2007}
\begin{equation}\label{fnlse-7}
\partial_{\tau}\psi(x,\tau)|_{\tau=0}=
\lim_{\tau\to 0}\frac{\psi(x,\tau) - \psi(x, 0)}{\tau}
= -C_{n,\alpha}(-\Delta_x)^{\frac{\alpha}{2}}f(x)\, .
\end{equation}

\subsubsection{Fractional NLSE in random potential}\label{sec:rp}

Now we account the effective potential $V$ in Eq. \eqref{gfse-2}
in the form of a random  potential, which describes a space disorder:
$V=V(x)$. Then the fractional NLSE \eqref{fnlse-1} reads
\begin{multline}\label{fnlse-8}
i\hh\partial_t\psi(x,t)
=(\hh^{\alpha}/2)(-\Delta_x)^{\frac{\alpha}{2}}\psi(x,t)
+\clB|\psi(x,t)|^2\psi(x,t)+V(x)\psi(x,t) \\
\equiv \hH(\alpha)\psi+\clB|\psi|^2\psi\, .
\end{multline}
Here $(-\Delta_x)^{\frac{\alpha}{2}}\equiv \clD_{|x|}^{\alpha}$
is defined in $\mathbb{R}$. For
the random potential $V=V(x), ~x\in(-\infty,+\infty)$,
Anderson localization takes place for both $\alpha=2$
\cite{Anderson,LGP} and $\alpha<2$ \cite{PaKoetal2020}
for the linear case $(\clB=0)$.
For $\alpha=2$, the situation is well studied and as is well known,
the system is described by the exponentially localized
Anderson modes (AM)s:
$\hH(\alpha=2)\Psi_{\omega_k}(x)=\omega_k\Psi_{\omega_k}(x)$,
where $\Psi_{\omega_k}\equiv\Psi_k(x)$ are
real functions and the eigenspectrum $\omega_k$ is discrete and
dense \cite{LGP}.
This problem of Eq. \eqref{fnlse-8} is relevant to experiments
in nonlinear optics, for
example disordered photonic lattices \cite{Moti2,lahini}, where
Anderson localization was found in the presence of nonlinear
effects ($\alpha=2$), as well as it relates to experiments on
Bose-Einstein condensates in
disordered optical lattices (see \textit{e.g.}, Refs.
\cite{SanchAspect,aspect}).
A typical example of fractional dynamics in optics is realized in
a competition between localization and nonlinearity that leads to
anomalous transport with a transport exponent $1/3$ observed numerically
\cite{fks} and analytically \cite{iom2010,MiIo2012}
(see also discussions in Refs. \cite{basko,SeMaSk2018}
and references therein).

Since the Hamiltonian $\hH(\alpha)$ is a self-adjoint operator
for all $\alpha\in (0,2]$,
the AMs $\Psi_k^{\alpha}(x)\equiv \Psi_k(x)$ are the complete set of
eigenfunctions (not obligatory real). Here we also suppose that the
random potential is large enough that the spectrum is
discrete\footnote{Delocalized states have not been observed yet,
and this issue stands for more detailed computational exploration
in the framework of a discrete fractional Anderson model
\cite{PaKoetal2020}.\label{sec:foot-a}}.
Then one projects Eq. \eqref{fnlse-8} on the basis of the AMs
\begin{equation}\label{fnlse-am2}
 \psi(x,t)=\sum_{\omega_k}C_{\omega_k}(t)
\Psi_{\omega_k}(x)=\sum_kC_k(t)\Psi_k(x)\,
 \end{equation}
and obtains a system of equations for coefficients of the expansion,
$C_k$
\begin{equation}\label{fnlse-am3}
 i\prt_t{C}_k=\omega_kC_k+
\clB\sum_{k_1,k_2,k_3}A_{k,k_1,k_2,k_3}
C_{k_1}^*C_{k_2}C_{k_3}\, ,
\end{equation}
where $ A_{k,k_1,k_2,k_3}$  is an overlapping integral of four AMs:
\begin{equation}\label{fnlse-am5}
 A_{k,k_1,k_2,k_3}
=\int\Psi_k^*(x)\Psi_{k_1}(x)\Psi_{k_2}^*(x)\Psi_{k_3}(x)dx\, .
 \end{equation}
The initial condition for system of Eqs.
(\ref{fnlse-am3}) is
$\psi_0(x)=\sum_ka_k\Psi_k(x)$. Equations (\ref{fnlse-am3})
correspond to a system of interacting classical nonlinear
oscillators with the Hamiltonian
\begin{equation}\label{fnlse-am4}
 H_{\rm osc}=\sum_k\omega_kC_k^*C_k+\frac{\clB}{2}
 \sum_{\bf k}A_{\bf k}
 C_{k_1}^*C_{k_4}^*C_{k_2}C_{k_3}\, ,
\end{equation}
where $\mathbf{k}=(k_1,k_2,k_3,k_4)$.

For $\alpha=2$, only the nearest neighbor oscillators are effectively
interacted \cite{MiIo2012,PaKoetal2020}, and the interaction
part of the Hamiltonian \eqref{fnlse-am4} is
$$H_{\rm int}=
\clB\sum_{k}A_{k,k,k,k}^{\pm} C_{k}^*C_{k-1}^*C_{k}C_{k+1}\, .$$
Then the transport in the chain of the nonlinear oscillators
is relevant to subdiffusion on a fractal Cayley
tree with the transport exponent being $1/3$ \cite{MiIo2012}.

The situation changes dramatically for Anderson localization of the L\'evy
flights with $\alpha<2$. As it follows from section
\ref{sec:em}, the Anderson Hamiltonian $\hH(\alpha)$ acts in the two
dimensional space that leads to the effective size of the
AMs $\Psi_k(x)$ being essentially extended \cite{PaKoetal2020},
and the number of effectively interacting oscillators can be large.
This large scale interaction resembles a quantum
relaxation process \cite{iom2016},
where the transition rate is determined by the Fermi's
golden rule \cite{schiff}.
Therefore, we follow the semiclassical qualitative estimation
according to Ref. \cite{MiIo2017} as follows.

Each nonlinear oscillator with the individual Hamiltonian
\begin{equation}\label{h_k}
h_k = \omega_k C^*_kC_k+\tfrac{\clB}{2}A_{k} C_{k}^*C_{k}^*C_{k}C_{k}
\end{equation}
represents one nonlinear eigenstate in the system, identified by
its wave number $k$, unperturbed frequency $\omega_k$, and nonlinear
frequency shift $\Delta\omega_k = \clB A_k C^*_kC_k$, where
$A_k\equiv A_{k,k,k,k}$.
Non-diagonal elements $A_{k,k_1,k_2,k_3}$
characterize couplings between each of the four
eigenstates with the wave numbers $k,k_1,k_2$, and $k_3$. It is
understood that the excitation of each eigenstate results
from the spreading of the wave field in the wave number
space. If the field spreads across a large number of states
$\Delta k\gg 1$, then the conservation of the probability
$$\int|\psi(x,t)|^2dx \sim \sum |C_k|^2\sim
\int |C_k|^2d\Delta k =1$$
implies that $|C_k|^2\sim 1/\Delta k$ and $|\psi(x)|^2\sim 1/\Delta x$.
Note that for the localized states
$1/\Delta x \sim 1/\Delta k $ \cite{iom2010}.
In the basis of AMs,
the evolution of the amplitudes $C_k$ from Eq. \eqref{fnlse-am3}
is controlled by the cubic nonlinearity:
$\dot{C}_k\sim\clB C_{k_1}^*C_{k_2}C_{k_3}$ .
The rate of excitation of the newly involved modes according to the
Fermi's golden rule is of the order
$\frac{d{|\psi}|^2}{dt} $
and proportional to the cubic power of the probability density,
$|\psi|^2$.
Taking the conservation of the probability into account,
one obtains that the rate of excitation is of the order
$\sim (1/\Delta k)^3$. On the other hand, the number of the newly
excited modes per unit time is $d\Delta k/dt$, making it possible
to assess $d\Delta k/dt \sim 1/(\Delta k)^3$.
This eventually yields subdiffusion with the mean squared displacement
$(\Delta x)^2\sim (\Delta k)^2 \propto t^{\half}$.
This essential increasing of the transport exponent from $1/3$ (when
$\alpha=2$) to $1/2$ results from the fractional Anderson Hamiltonian
$\hH(\alpha)$ for $\alpha<2$. Note that this result for the
transport exponent is
corroborated to an experimental observation of the optically
induced exciton transport in molecular crystals, which exhibits
the intermediate asymptotic subdiffusion \cite{Aksel_etal2014} with
the experimental transport exponent of the order of $\sim 0.57$.

\subsection{L\'evy flights in Schr\"odinger - Newton
equation}\label{sec:sne}

It has been shown that gravitational effects can be studied
in optical experimental setup \cite{Moti1}, which can emulate
general relativity, gravitational effects with optical wave
packets under a long-range nonlocal thermal nonlinearity.
The Helmholtz equation, which describes this experiment
in paraxial regime,
is reduced to the Schr\"odinger - Newton equation (SNE)
\cite{Moti1}.
The latter has been proposed to describe the gravitational
self-interaction of quantum wave packets.
In the form of coupling, the
Schr\"odinger equation of quantum mechanics with the
Poisson equation from Newtonian mechanics, the SNE reads
\begin{subequations}\label{sne-1}
\begin{align}
i\hh\partial_t\psi=-\frac{\hh^2}{2m}\Delta\psi+U\psi\, , \label{sne-1a}
\\
\Delta U=4\pi Gm^2|\psi|^2\, . \label{sne-1b}
\end{align}
\end{subequations}
Here $G$ is Newton's constant, $m$ is the mass of a quantum
particle in the gravitational potential, which is
determined by the density of the wave function $\psi=\psi(\mathbf{r},t)$.

This Schr\"odinger - Newton equations has been introduced by Penrose
in connection with the role of gravity in quantum state reduction
(collapse of wave functions) cite{Penrose1}.
Another form of the SNE, introduced by Di\'osi \cite{Diosi} and
Penrose \cite{Penrose2} reads
\begin{equation}\label{sne-2}
i\hh\partial_t\psi=\left[-\frac{\hh^2}{2m}\Delta-
Gm^2\int\frac{|\psi(\mathbf{r}')|^2}{|\mathbf{r}-\mathbf{r}'|}d^3\mathbf{r}'
\right]\psi\, .
\end{equation}
Since the seminal results, these equations have been extensively studied,
see Refs. \cite{Penrose3,Bahrami2014} and references therein.

The fractional generalization of the Poisson Eq. \eqref{sne-1b}
for Newtonian gravity has been suggested recently as a novel approach
for Galactic dynamics, considered in fractional Newtonian gravity,
see discussion in Ref. \cite{guisti2020}. Following this
consideration of a fractional gravitational potential, we
consider the fractional Laplacian for both Eqs. \eqref{sne-1a} and
\eqref{sne-1b}.
To give a brief insight into the stability of the dynamics with respect
to four L\'evy waves processes, we simplify the consideration
by reducing it to the one dimensional (1D) case
of Eq. \eqref{sne-2}.
Therefore, the 1D fractional SNE in the form of Eq.
\eqref{gfse-2} reads as follows
\begin{equation}\label{sne-3}
i\hh\partial_t\psi=\left[\frac{\hh^{\alpha}}{2m}(-\Delta_x)^{\alpha/2}-
\frac{Gm^2}{\Gamma(-\nu)\cos\left(\frac{\nu\pi}{2}\right)}
\int\frac{|\psi(x')|^2}{|x-x'|^{1+\nu}}dx'\right]\psi\, ,
\end{equation}
where $\nu\in (0,1)$. The last term in Eq. \eqref{sne-3} is
just the Riesz fractional derivative of the density of the wave function.
The zero boundary conditions are taken at infinity.
The initial conditions will be specified when necessary later.
Performing the Fourier transformation of the equation and taking into
account the Fourier presentation of the wave function
\begin{equation}\label{sne-4}
\psi(x,t)=\frac{1}{2\pi}\int A(k,t) e^{ikx}dk\, ,
\end{equation}
we have
\begin{equation}\label{sne-5}
i\hh\dot{A}_k=\frac{\hh^{\alpha}}{2m}|k|^{\alpha}A_k
-Gm^2
|k|^{\nu}\int dk_1dk_2dk_3 A_{k_1}^{*}A_{k_2}A_{k_3}
\delta(k_2+k_3-k-k_1)\, ,
\end{equation}
where $A_k(t)\equiv A(k,t)$.
The initial conditions are such that at the moment $t=0$
the only mode with fixed $k=q$ is populated, then
\begin{equation}\label{sne-6}
A_k(t=0)=a_k\delta(k-q)\, .
\end{equation}
The solution of Eq. \eqref{sne-5} with the initial condition
\eqref{sne-6} reads
\begin{equation}\label{sne-7}
A(k,t)=\left\{\begin{split}
& \exp\big[-i\frac{\hh^{\alpha-1}}{2m}|q|^{\alpha}t
+\frac{Gm^2}{\hh}|q|^{\nu}|a_q|^2t \big]a_q  &\text{ if $k=q$} \, , \\
& 0 & \text{if $k\neq q$}\, .
\end{split}\right.
\end{equation}
In this case, the evolution of the wave function is due to the
solution in Eq. \eqref{sne-7}, namely $\psi(x,t)=A(q,t)e^{iqx}$.

\subsubsection{Stability analysis}

Let us investigate stability of the solution \eqref{sne-7}
with respect to decay in the nearest modes according to the resonant
condition $2q\rightarrow (q+p)+(q-p)$. In this case, at the initial moment
all other modes with $k\neq q$ are populated with amplitudes
$|a_k|\ll |a_q|$. In fact, the $k$'s amplitudes can be
considered infinitesimally
small. Then a perturbation theory can be applied to Eq. \eqref{sne-5}.
Taking into account only linear terms with respect to amplitudes
$A_{q\pm p}$ and $A^*_{q\pm p}$, according to
Refs.~\cite{Preprint,BeKo84},
the closed system of equations
reads
\begin{subequations}\label{sne-sa1}
\begin{align}
i\hh\dot{A}_q &=\frac{\hh^{\alpha}}{2m}|q|^{\alpha}A_q
-Gm^2|q|^{\nu}|A_q|^2A_q \label{sne-sa1a} \\
i\hh\dot{A}_{q+p}& =\frac{\hh^{\alpha}}{2m}|q+p|^{\alpha}A_{q+p}
-2Gm^2|q+p|^{\nu} |A_q|^2A_{q+p}-Gm^2|q+p|^{\nu} A_q^2A^*_{q-p} \nonumber \\
i\hh\dot{A}_{q-p}& =\frac{\hh^{\alpha}}{2m}|q-p|^{\alpha}A_{q-p}
-2Gm^2|q-p|^{\nu} |A_q|^2A_{q-p}-
Gm^2|q-p|^{\nu} A_q^2A^*_{q+p} \, . \label{sne-sa1b}
\end{align}
\end{subequations}
It follows from Eqs. \eqref{sne-sa1a} that the dynamics of the
``large'' wave \eqref{sne-7} is not changed in the first order of the
perturbation theory.
In contrast,
amplitudes of the ``small'' waves grow exponentially.
To show this, we simplify the form of Eqs \eqref{sne-sa1}
by introducing new variables:
\begin{align*}
& F=\frac{\hh^{\alpha-1}}{2m}|q|^{\alpha}
+\frac{Gm^2}{\hh}|q|^{\nu}I\equiv \omega(q)+\Omega(q)I\, , \\
& F(\pm)=\frac{\hh^{\alpha-1}}{2m}|q\pm p|^{\alpha}
-2\frac{Gm^2}{\hh}|q\pm p|^{\nu}I\equiv \omega(\pm)-\Omega(\pm)I\, , \\
&\clF=F-[F(+)+F(-)]/2\, , \quad |A_q(t)|^2=|a_q|^2=I\, , \\
&A_{q\pm p}(t)\equiv A_{\pm}(t)=\exp[-i(F+\clF)t]C_{\pm} \, .
\end{align*}
Then the exponential growth with the increment $\lambda_p$ is determined
by equation
\begin{equation}\label{sne-sa2}
\begin{pmatrix}
\dot{C}_{+} \\ \dot{C}_{-}\end{pmatrix}
=
\begin{pmatrix}
i\clF & i\Omega(+)I \\ -i\Omega(-)I & -i\clF \end{pmatrix}
\begin{pmatrix} C_{+} \\ C_{-}\end{pmatrix} \, .
\end{equation}
Solving the characteristic equation of the matrix
and performing expansion of the expression
$|q+p|^{\alpha}+|q-p|^{\alpha}-2|q|^{\alpha}
\approx \alpha(\alpha-1)|q|^{\alpha-2}p^2/2$, where
$p\ll q$ and $|q|^{\alpha-1}\delta(q)=0$ since $q\neq 0$,
one obtains the increment of the order of
$$\lambda_p\sim |q|^{\nu}Gm^2
\sqrt{I^4 -|q|^{-\alpha-\nu}+\mathrm{O}(p^4)}\, .$$
Since, according to the normalization of the wave function
$|a_q|\sim 1$, while $q>1$, the real part of increment is always
positive value.
Therefore, if initially only few modes are populated, these modes
are always unstable with respect to the four L\'evy waves decay.
In opposite case, when the initial condition corresponds to
thermalisation and all $k$ modes are equally populated with
$|a_k|\rightarrow 0$. Then the increment reads
$$\lambda_p\sim \pm i
p^2\frac{\hh^{\alpha-1}}{4m}|(1-\alpha)q^{\alpha-2}|\alpha$$
and is imaginary. Then all the modes are stable.
In this case, the gravitational, nonlinear term is infinitesimally
small and can be neglected. Then the wave function can be
estimated in the terms of the Fox $H$-function,
\begin{multline}\label{sne-sa3}
\psi(x,t)=\frac{\clN}{2\pi N}\int^{N/2}_{-N/2} \exp\left[ikx
-i\frac{\hh^{\alpha-1}}{2m}|k|^{\alpha}t\right]dk \\
\approx\frac{\clN}{2\pi N}\int_0^{\infty} \cos(kx)
H_{0,1}^{1,0}\left[i\frac{\hh^{\alpha-1}}{2m}k^{\alpha}t
\left|\begin{array}{cc} - \\ (0,1)\end{array}\right.\right]dk \\
=\frac{1}{x}\frac{\clN}{2 N}
H_{2,2}^{1,1}\left[-i\frac{2m}{t\hh^{\alpha-1}}|x|^{\alpha}
\left|\begin{array}{cc} (1,1),(1,\frac{\alpha}{2})\\ (1,\alpha),(1,0),(1,\frac{\alpha}{2})\end{array}\right.\right]\, .
\end{multline}
Here, $N\rightarrow \infty$ is a number of modes, and eventually
it will be absorbed by normalization constant $\clN=2\pi N$.

\section{Fractional time Heisenberg equation vs FTSE}
\label{sec:part2}

When parameter $\beta\neq 1$ in Eq. \eqref{gfse-2},
the fractional time quantum mechanics
is not Markovian \cite{Ta08,La2017,iom2019}
and nonunitary:
it violates the Stone's theorem on the
one-parameter unitary group \cite{stone}.
The Feynman path integral approach
does not exist,
and as a result, the equivalence between
Schr\"odinger and Heisenberg pictures of the quantum mechanics
is broken \cite{iom2019,ErDeSiBu10}.
However, the fractional Heisenberg
equation of motion (FHEM) can be obtained by
standard quantization of classical
equations of motion with memory \cite{iom2019}
i.e. with fractional time derivative of Eq. \eqref{int-1}.
For better understanding some arbitrariness in the formulation
of the problem, we first, obtain a FTSE
\begin{equation}\label{ftse-1}
i\hh\partial^{\beta}_t\psi(x,t) =\hat{H}\psi(x,t)\, ,
\end{equation}
where $\hat{H}$ is the Hamiltonian operator for $\beta=1$
only\footnote{When $\beta\neq 1$ the system contains also a specific
relaxation and cannot be separated from the environment.
Note that independently of $\beta$, operator $\hat{H}$ is a well defined
Hermitian operator, and in sequel, we consider it as the
Hamiltonian operator.}. Following
Refs. \cite{AcYaHa13,iom2019}, we consider the FTSE \eqref{ftse-1}
suggesting some additional arguments of its inferring.

\subsection{Properties of the FTSE: Infinitesimal evolution}\label{sec:ftse}

In Ref. \cite{AcYaHa13}, the FTSE \eqref{ftse-1} was obtained
as a quantisation of the classical fractional dynamics in the framework
of the path integral representation. However, this approach
supposes the chain rule in the form of Eq. \eqref{Smoluch}
together with the unitary dynamics of the Green's functions.
Although this approach looks controversial, it is not so
if we avoid a construction of any path integral and
consider the fractional dynamics on the infinitely small
time scale $\Delta t \rightarrow 0$ that eventually has been performed
in Ref.  \cite{AcYaHa13} and then in Ref. \cite{iom2019}.
As is well known, the fractional evolution
according to the FTSE is due to the Mittag-Leffler function,
which reads from Eqs. \eqref{ftse-1} and \eqref{A9a} as follows
\begin{equation}\label{ftse-2}
\psi(t+\Delta t)=E_{\beta}
\left(-i\hat{H}\Delta t^{\beta}/\hbar\right)\psi(t) \, .
\end{equation}
For the small argument it yields the stretched exponential behavior
of the evolution operator on the infinitesimal time scale $\Delta t$
\begin{multline}\label{ftse-3}
\psi(t+\Delta t)=\hat{U}\left(t+\Delta t,t\right)\psi(t)=
\exp\left[\tfrac{-i\hat{H}\Delta t^{\beta}}{\hbar\Gamma(\beta+1)}\right]
\psi(t) =\\
=\exp\left[\frac{-i}{\hh\Gamma(\beta)}
\int_t^{t+\Delta t}(t+\Delta t-t')^{\beta-1}\hat{H}dt'\right]
\psi(t) \, .
\end{multline}
Note, that for $\Delta t\rightarrow 0$, we perform fractional
Taylor expansion, defined in Eqs. \eqref{A12} and \eqref{A13},
for the l.h.s. of Eq. \eqref{ftse-3} and
Taylor expansion for the r.h.s. of the equation.
This immediately yields the FTSE \eqref{ftse-1}.
Note that fractional Taylor expansion for the l.h.s.
of Eq. \eqref{ftse-3} destroys the Markov chain rule.

Now we can obtain the FTSE by means of the
Green's function on an infinitesimal time scale.
As it has been shown above, the exponential form of the
infinitesimal Green's function can be used, and
the infinitesimal evolution of the wave
function reads
\begin{equation}\label{ftse-4}
\psi(x,t+\vep)\equiv \langle x|\psi(t+\vep)\rangle=
\int_{-\infty}^{\infty}\clG(x,t+\vep|y,t)\psi(y,t)dy,
\end{equation}
where the infinitesimal Green's function is determined by  conjecture
\begin{equation}\label{ftse-5}
\clG(x,t+\vep|y,t)=\frac{1}{A}
\exp\left[iS(\vep,x,y)/\hh\right]\, ,
\end{equation}
according to the principal Hamilton function, or action $S(\vep,x,y)$.
The latter is defined by the functional
\begin{equation}\label{ftse-6}
S[f]=\frac{1}{\Gamma(\beta)}\int_t^{t+\vep}(t+\vep-\tau)^{\beta-1}
L\left(\partial^{\beta}_{\tau}f(\tau),f(\tau)\right)d\tau
\end{equation}
with the boundary conditions $f(t)=y$ and $f(t+\vep)=x$.
The ``fractional'' action plays the role of a
generating functional of the quantum
evolution on the infinitesimal time scale, and as it follows
from the fractional evolution operator \eqref{ftse-3},
it has ``fractional time metric''.
The Lagrangian $L=\tfrac{(\partial_t^{\beta}f)^2}{2}-V(f)$
corresponds to the
classical counterpart of the FTSE with the Hamiltonian $\hat{H}$.
On the infinitesimal time scale $\vep<\Delta t\rightarrow 0$
the potential $V(f)$ contributes to the action
$S(\vep,x,y)$ according to the middle
point \cite{feynman}, which yields
$-V\left(\tfrac{x+y}{2}\right)\vep^{\beta}/\Gamma(\beta+1)$.
Therefore, we find the fractional classical dynamics due to the
kinetic part, which is the fractional integral
$$S_{\rm e}[f]=\frac{1}{\Gamma(\beta)}
\int_t^{t+\vep}(t+\vep-\tau)^{\beta-1}
L\left(\partial^{\beta}_{\tau}f(\tau)\right) d\tau\, , $$
obtained in \ref{sec:app_B}. Taking into account
Eq. \eqref{B-pi-13} for the Green's function \eqref{ftse-5},
we obtain the infinitesimal evolution of the wave function
in the Feynman's form (see chapter 4 in Ref.~\cite{feynman}),
which reads
\begin{multline}\label{ftse-7}
\psi(x,t+\vep)=\frac{1}{\mathbb{A}}\int_{-\infty}^{\infty}
\exp\left[\frac{i}{\hbar\Gamma(\beta+1)}
\frac{m_{\beta}(x-y)^2}{2\vep^{\beta}}\right] \\
\times\exp\left[-\frac{i\vep^{\beta}}{\hh\Gamma(\beta+1)}
V\left(\tfrac{x+y}{2}\right)\right]\psi(y,t)dy\, .
\end{multline}
Here $$\mathbb{A}=\left(\frac{2\pi i\hh\vep^{\beta}}
{m_{\beta}/\Gamma(\beta+1)} \right)^{\frac{1}{2}}$$
is defined in Eq. \eqref{B-pi-13b} and
$m_{\beta}=\Gamma^2(\beta+1)$ is an effective
dimensionless mass due to the fractional time dynamics.
Then introducing the infinitesimal shift $\eta=y-x$ and
following Feynman's instruction, we make the
change of the variable $dy=d\eta$ then perform expansion
and integration with respect to $\eta$. These procedures result
in the r.h.s. of Eq. \eqref{ftse-7} as follows
\begin{equation}\label{ftse-7-8}
\psi(x,t+\vep)=\psi(x,t)+\frac{\vep^{\beta}}{\Gamma(\beta+1)}
\left[-\frac{\hh^2}{2m_{\beta}}\partial^2_x+V(x)\right]\psi(x,t)\, .
\end{equation}
To relate the both sides of the equation with the same
power $\vep^{\beta}$, we perform fractional Taylor
expansion, which yields according to Eq. \eqref{A13}
\begin{equation}\label{ftse-8}
\psi(x,t+\vep)=\psi(x,t)+\frac{\vep^{\beta}}{\Gamma(\beta+1)}
\partial^{\beta}_t\psi(x,t) + \mathrm{O}(\vep^{\beta+1})\, .
\end{equation}
Eventually, taking the limit $\vep=0$, we obtain the FTSE in Eq.
\eqref{ftse-1} with the Hamiltonian operator
$\hat{H}=-\tfrac{\hh^2}{2m_{\beta}}\partial^2_x+V(x)$.

\subsection{Symmetrical form of FHEM}\label{sec:sf}

Since the fractional evolution is not unitary,
the equivalence between
Schr\"odinger and Heisenberg representations
of fractional time quantum mechanics is broken.
In this case, we have some arbitrariness in formulation of the FHEM.
A possible convenient
way of construction of the FHEM by analogy with
the FTSE \eqref{ftse-1}, is replacing time derivative
$d/dt$ by fractional time derivative $\partial^{\alpha}_t$
in the Hamiltonian structure, known as the Poisson brackets
\[\{f,g\}=\frac{\partial f}{\partial p}\frac{\partial g}{\partial x}
-\frac{\partial f}{\partial x}\frac{\partial g}{\partial p}\]
(with the classical counterpart of the Hamiltonian $\hH$).
Here $f,g$ are arbitrary functions of
the momentum $p$ and  the coordinate $x$.
Therefore the fractional evolution of an arbitrary function
$f(x,p)$ is determined by the Poisson brackets
\begin{equation}\label{fhem-3}
\mathcal{K}f(p,q)\equiv \{H,f\}=
\left(\frac{\partial H}{\partial p}\frac{\partial}{\partial x}-
\frac{\partial H}{\partial x}\frac{\partial}{\partial p}\right)
f(p,q)\, ,
\end{equation}
and the fractional classical dynamics is described by fractional
equations of motion
\begin{equation}\label{fhem-4}
\partial^{\beta}_tf(x,p)=\clK f(x,p)\, .
\end{equation}
Here $x$ and $p$ are taken at some fixed time, for example at
$t=0$.

Now we return to our main consideration of the FHEM by quantization of
the Poisson brackets
$\{x,p\}\rightarrow [\hat{x},\hat{p}]=-i\hh\{x,p\}=i\hh$.
Therefore, the FHEM for an arbitrary operator
$\hat{f}=\hat{f}\left(\hat{x}(t),\hat{p}(t)\right)$ reads
\begin{equation}\label{fhem-5}
-i\hh\partial^{\beta}_t\hat{f}=[\hat{H},\hat{f}]\, ,
\end{equation}
where the initial condition of the operator is
$\hat{f}_0=\hat{f}_0\left(\hat{x},\hat{p}\right)$ and
its evolution is the explicit function of time.
Note that $\left(\hat{x},\hat{p}\right)$ are taken at $t=0$.

\subsubsection{Oscillator dynamics in coherent states}\label{sec:cs}

An important property of Eq. \eqref{fhem-5} is
that the dynamics is not unitary and as a
consequence, the commutator
$[\hat{f},\hat{H}]\neq \hat{U}^{\dag}(t)[\hat{f}_0,\hat{H}_0]\hat{U}(t)$
is not defined at arbitrary time $t$.
However, the Hamiltonian is the integral
of motion since $[\hat{H},\hat{H}]=0\, , ~\forall\, t$ and
for conservative systems this yields
$\hat{H}\left(\hat{x}(t),\hat{p}(t),t\right)=\hat{H}(t=0)=
\hat{H}_0(\hat{x},\hat{p})$.
In this case, it is possible to consider the commutator at $t=0$.

To explain this situation, we consider a system described
in the framework of the Bose creation, $\had$ and
annihilation, $\ha$ operators, which are taken at $t=0$
with the commutation rule $[\ha,\had]=1$ and
the Hamiltonian $\hat{H}=\hat{H}(\hat{a}^{\dag},\ha)$.
Here $\ha=\ha(t=0)$ and $\had=\had(t=0)$.
We can also rewrite the operator $\hat{f}$ as a function of
the creation and annihilation operators
$\hat{f}(t)=\hat{f}\left(\had,\ha,t\right)$.

Now we can introduce a basis of coherent states $|a\rangle\, ,
\langle a|$ at $t=0$ as the eigenfunctions\footnote{It also relates
to a specific superposition of the Fock states of a harmonics oscillator.
However, for our purpose, Eq. \eqref{bcs-1} is a satisfactory definition.}
of the creation and annihilation operators $\had$ and $\ha$ at $t=0$,
respectively:
\begin{equation}\label{bcs-1}
\ha|a\rangle=a|a\rangle\, , ~ ~~\langle a|\had=a^*\langle a|\, .
\end{equation}
Using this basis, we introduce the average value of the operator function
\begin{equation}\label{bcs-2}
f(t)\equiv f(a^*,a,t)=\langle \hat{f}(t)\rangle=
\langle a|\hat{f}\left(\had,\ha,t\right)|a\rangle\, .
\end{equation}
This formulation supposes the normal ordering
of the operator with respect to the
creation and annihilation operators: $\hat{f}=
\sum_{m,n}f_{m,n}(t)(\had)^m\ha^n$.
The mapping procedure of FHEM \eqref{fhem-5}
on the basis of the coherent states is based on the
following properties \cite{SiTs82}
\begin{subequations}\label{bcs-3}
\begin{align}
\langle a|\hat{f}\left(\had,\ha,t\right)\had|a\rangle
=e^{-|a|^2}\frac{\partial}{\partial\, a}e^{|a|^2}f(a^*,a,t)\, ,
\label{bcs-3a} \\
\langle a|\ha\hat{f}\left(\had,\ha,t\right)|a\rangle
=e^{-|a|^2}\frac{\partial}{\partial\, a^*}e^{|a|^2}f(a^*,a,t)\, .
\label{bcs-3b}
\end{align}
\end{subequations}
The averaging procedure results in the $c$-number (scalar valued)
fractional equation of motion as follows
\begin{equation}\label{bcs-4}
\partial_t^{\beta}f(t)=\frac{i}{\hh}\hat{\clK}f(t)\, ,
\end{equation}
where $\hat{\clK}$ is a quantum Koopman operator, or
$\clK$-operator, which reads \cite{iom2016,SiTs82,BeIo85,iom17}
\begin{equation}\label{bcs-5}
\hat{\clK}=e^{-|a|^2}\left[
H\left(a^*\, ,\tfrac{\partial}{\partial a^*}\right)
-H\left(\tfrac{\partial}{\partial a} ,a\right)\right]e^{|a|^2}\, .
\end{equation}

\subsubsection{Hamiltonian form of FHEM}\label{sec:Hf}

As shown above, the symmetrical form of the FHEM yields a simple
and elegant way to find the quantum evolution in the framework of the
Koopman operator \eqref{bcs-5}. However, it is tempting
to obtain the FHEM by quantization of the Hamilton
equation of motion, obtained from the fraction Lagrangian
$$L=\frac{1}{2}
\left(\partial_t^{\beta}x\right)^2-V(x)\equiv v_{\beta}^2-V\, .$$
Then we obtain the momentum $p_{\beta}=(\partial L/\partial v_{\beta})$,
and the Hamiltonian is
\begin{equation}\label{fhem-1}
H=p_{\beta}v_{\beta}-L=\frac{1}{2}p_{\beta}^2+V\, .
\end{equation}
Therefore, fractional Hamilton equations of motion,
defined by the Lagrange-Euler equations are ($p_{\beta}\equiv p$)
\begin{subequations}\label{fhem-2}
\begin{align}
\partial_t^{\beta}x  =p \label{fhem-2a} \\
\fourIdx{RL}{t}{\beta}{T}{\clD}p  =-\frac{\partial V(x)}{\partial\, x} \, .
\label{fhem-2b}
\end{align}
\end{subequations}
The first equation \eqref{fhem-2a} with the Caputo fractional
derivative corresponds to the
symmetrical $c$-number fractional equation of motion \eqref{bcs-4},
which eventually yields a decay solution, when $p=p(t)$ is known.
The second equation
\eqref{fhem-2b} with right Riemann-Liouville fractional derivative
for the averaged momentum
$p(t)\equiv p(a^*,a,t)=\langle a|\hat{p}|a\rangle$
yields
\begin{equation}\label{bcs-hf-1}
\fourIdx{RL}{t}{\beta}{T}{\clD} p(t)=\frac{i}{\hh}
\hat{\clK} p(t)\, ,
\end{equation}
where we used that $-\frac{\partial V(x)}{\partial\, x}\equiv \clK p$.
Solving this equation, we first pay attention at the
particular solution $p(t)=C(T-t)^{\beta-1}$.
Taking into account that
$\fourIdx{RL}{t}{\nu}{T}{D}(T-t)^{\nu-1}=0$, and
the momentum is invariant with respect to the time shift
$C(T-t)^{\beta-1}$, where $C$ is such that
 $C=C(a^*,a)$ is the
invariant\footnote{For example, for the oscillators
we have $C=|a|^2$. Another possibility
is a ``Hamiltonian'' $C=H(a^*,a)$, since the condition
$\hat{\clK}H(a^*,a)=0$ is always fulfilled.} of the motion:
$\hat{\clK}C=0$, then Eq. \eqref{bcs-hf-2} is a particular solution.
Therefore, looking for the momentum
as a function of $T-t$, and taking into account the variable
separation $p(t)=\clU(t)X(a,a^*)$, then the momentum in
Eq. \eqref{bcs-hf-1} is the expansion
\begin{equation}\label{bcs-hf-2}
p(t)=\sum_{n=0}^{\infty}c_{n,\beta}(T-t)^{n\beta}X(a,a^*)\, .
\end{equation}
Substituting Eq. \eqref{bcs-hf-2} in Eq. \eqref{bcs-hf-1}, one obtains
\begin{equation}\label{bcs-hf-3}
p(t)=\clU(t)X(a,a^*)=
E_{\beta}\left[i(T-t)^{\beta}\hat{\clK}/\hh\right]X(a,a^*)\, .
\end{equation}
Here $X(a,a^*)$ is a stationary part of the momentum. For example,
it can be considered as the ``initial condition'' at $t=T$.
Then from Eq. \eqref{bcs-hf-3}, it follows $p(t=T)=p(0)=X(a,a^*)$.
Another possibility, it can satisfy the eigenvalue equation,
$\hat{\clK} X(a,a^*)=\hh\lambda X(a,a^*)$.
In this case, the nonunitary dynamics of the momentum is due
to the Mittag-Leffler function
$\clU(t)=E_{\beta}\left[i\lambda(T-t)^{\beta}\right]$.

\section{Conclusion}

In the review, we have concerned with few examples of
fractional Schr\"odinger equations (FSE)s as descriptive
models of light propagation in nonlinear composite
materials envisaged for real experimental tasks.
The general form of the FSE  \eqref{gfse-2} describes two large
classes of non-local quantum mechanics, namely  fractional space
Schr\"odinger equations (FSSE)s and fractional time Schr\"odinger equations
(FTSE)s considered in the review. These two classes belong to completely
different quantum phenomena. The former one corresponds to quantum L\'evy
walks, including Anderson localization of the L\'evy flights, which are
described by the Hamiltonian quantum mechanics.
In contrast, the FTSEs describe non-Markovian quantum dynamics,
including quantum
decay/friction due to interactions with the environment. However, in optical
context, when time is considered as an effective presentation
of a longitudinal coordinate in the parabolic equation approximation,
the FTSE relates
to linear - nonlinear optics in curved spaces and fractional
differential geometry. Unfortunately, the latter is a vague
issue in fractional calculus. Note also that ``fractional time metric''
$(T-t)^{\beta-1}$ is reasonable to use
only for the infinitesimal evolution for the generating action
for the fractional Lagrangian. In this connection,
the extension method with fractional metric looks as a
possible way to overcome this obstacle, see \textit{e.g.},
Ref.~\cite{zacher}.
The similar approach for the FTSE can be suggested in the framework of
the non-fractional Lagrangian with fractional metric
in the fractional action \eqref{ftse-6},
defined on the infinitesimal fractional time. In Sec.
\ref{sec:ftse}, this situation was discussed in the framework
of generating functional for the FTSE, where
a fractional time derivative in the classical
dynamics invokes fractional time quantum mechanics,
see also \ref{sec:app_B}.

In an alternative consideration of this fractional
time functional for arbitrary time $t$  with classical
Lagrangian $L_0=\frac{\dot{q}^2}{2}-V(q)$,
the fractional action reads
\begin{equation}\label{fau-1}
S[q]=\frac{1}{\Gamma(\nu)}\int_0^tL_0(\dot{q},q,\tau)
(t-\tau)^{\nu-1}d\tau\, ,
\end{equation}
where the variable $q$ is a coordinate in some $d$-dimensional space.
This kind of fractional actions, presented in the form of the
Lebesgue-Stieltjes integral is considered in quantum
gravity \cite{Calcagni}, where the standard measure
in the action is replaced by a nontrivial measure.
An example of such fractional action \eqref{fau-1}
has been introduced
in the Lagrangian dynamics with friction \cite{ElNab05}, as well.

The Lagrangian of a non-conservative system
\[L(\dot{q},q,\tau)=\frac{1}{\Gamma(\nu)}
L_0(\dot{q},q,\tau)(t-\tau)^{\nu-1}\, \]
produces the Lagrangian equations of motion
\begin{subequations}\label{fau-2}
\begin{align}
\frac{d\,L}{d\, q}-\frac{d}{d\tau}\frac{\partial L}{\partial\dot{q}}=0
\, , \label{fau-2a}\\
\ddot{q}+\partial_qV(q)+\frac{1-\nu}{t-\tau}\dot{q}=0
\, . \label{fau-2b}
\end{align}
\end{subequations}
In the new time, $t-\tau\rightarrow t$ with
the time derivative $d/dt$,  Eq.
\eqref{fau-2b} reads
\begin{equation}\label{fau-3}
\ddot{q}+\partial_qV(q)-\frac{1-\nu}{t}\dot{q}=0\, .
\end{equation}
Then the Lagrangian is $L=[\dot{q}^2/2-V(q)]t^{\nu-1}$,
and the corresponding action is not anymore fractional
$S[q]=\int_0^tL(\dot{q},q,\tau)d\tau$ as well.
The Hamiltonian dynamics can be constructed
as well. Introducing the momentum
$p=\frac{\partial L}{\partial \dot{q}}=\dot{q}t^{\nu-1}$,
we obtain the Hamiltonian
\begin{equation}\label{fau-4}
H(p,q,t)=\frac{p^2}{2}t^{1-\nu}+V(q)t^{\nu-1}\, ,
\end{equation}
and the Schr\"odinger equation reads
\begin{equation}\label{fau-5}
i\hbar\partial_t\psi(q,t)=\hat{H}(t)\psi(q,t)\, .
\end{equation}
It also follows from Eq. \eqref{fau-1} that $H(p,q,t)=
-\partial_tS[q]=-\clD_t^{1-\nu}L_0(\dot{q},q,\tau)$.
Therefore, the Hamiltonian (Markov) quantum mechanics
can be constructed to describe the fractional time dynamics.
Correspondingly, the Heisenberg equations of motion
can be constructed as well\footnote{Note also that the
path integral can be constructed as well, since
the quantum Hamiltonian \eqref{fau-4} produces the classical action
$S[q]$ rigorously by means of the Trotter product formula,
see \textit{e.g.} \cite{schulman}.}.
This issue needs further clarification with respect to
local differential geometry
in linear - nonlinear optics in curved space \cite{Rivka1}.

\appendix

\section{Elements of fractional integro--differentiation}\label{sec:app_A}

A basic introduction to fractional calculus can be found,
{\em e.g.}, in Refs. \cite{oldham,Po99}. Fractional integration of the
order of $\alpha$ is defined by the operator\footnote{Greek letters
$\alpha,\beta,\nu$ are arbitrary used for the order of the fractional integro-differentiation, while Latin letters $t$ and $x$ are used as  variables.}
\begin{equation}\label{A1}
\fourIdx{}{a}{\alpha}{t}{I}f(t)
=\frac{1}{\Gamma(\alpha)}
\int_a^tf(\tau)(t-\tau)^{\alpha-1}d\tau, ~~(\alpha>0)\, ,
\end{equation}
where $\Gamma(\alpha)$ is a gamma function, and there is no constraint
on the lower limit $a$. A fractional derivative is defined as
an inverse operator to fractional integration
$\fourIdx{}{a}{\alpha}{t}{I}$ in Eq. \eqref{A1}.
Its explicit form is the convolution
\begin{equation}\label{A2}
\fourIdx{}{a}{-\alpha}{t}{I}=
\fourIdx{}{a}{\alpha}{t}{\clD} =
\frac{1}{\Gamma(-\alpha)}\int_0^t
\frac{f(\tau)}{(t-\tau)^{\alpha+1}}d\tau \, .
\end{equation}
For arbitrary $\alpha>0$, this integral is, in general, divergent.
As a regularization of the divergent integral, the following two
alternative definitions for \fourIdx{}{a}{\alpha}{t}{\clD}  exist.
The first one is the Riemann--Liouville fractional derivative.
For $ n-1<\alpha<n,~~n=1,2,\dots$ it reads
\begin{equation}\label{A3}
\fourIdx{RL}{a}{\alpha}{t}{\clD}f(t)
=D_t^n \fourIdx{}{a}{n-\alpha}{t}{I}f(t) =
\frac{1}{\Gamma(n-\alpha)}\frac{d^n}{dt^n}\int_a^t
\frac{f(\tau)d\tau}{(t-\tau)^{\alpha+1-n}} \, .
\end{equation}
The second one
\begin{equation}\label{A4}
 \fourIdx{C}{a}{\alpha}{t}{\clD}f(t)=
 \fourIdx{}{a}{n-\alpha}{t}{I}D_t^nf(t) =
 \frac{1}{\Gamma(n-\alpha)}\int_0^t
\frac{f^{(n)}(\tau)d\tau}{(t-\tau)^{\alpha+1-n}} \, ,
\end{equation}
is the fractional derivative in the Caputo form.

For the time, where the limit is $a=0$, we use
$\fourIdx{C}{0}{\beta}{t}{\clD}f(t)\equiv \partial_t^{\beta}$
for the Caputo derivative and
$\fourIdx{RL}{0}{\beta}{t}{\clD}\equiv \clD_t^{\beta}$ for
the Riemann--Liouville derivative.

The Laplace transformation can be performed, as well.
For example for Eq. (\ref{A4}),
if $\clL[f(t)]=\tilde{f}(s)$ is the Laplace transformation, then
\begin{equation}\label{A8}  %
\clL\left[\partial_t^{\beta}f(t)\right]=s^{\alpha}\tilde{f}(s)-
\sum_{k=0}^{n-1}f^{(k)}(0^+)s^{\beta-1-k}\, .
\end{equation}

When $a=-\infty$, the resulting fractional derivative is the
Weyl derivative,
\begin{equation}\label{A8-1}
\mathcal{W}^{\nu}f(x)\equiv\fourIdx{}{-\infty}{\nu}{x}
{\mathcal{W}}f(x)\equiv
\fourIdx{W}{-\infty}{\nu}{x}{\mathcal{D}}f(x)=\frac{1}{\Gamma(-\nu)}
\int_{-\infty}^x\frac{f(x')dx'}{(x-x')^{1+\nu}}.
\end{equation}
If we impose the physically reasonable condition $f(-\infty)=0$
together with its $n$ derivatives, where $n-1<\nu<n$, then
\begin{equation}\label{A8-2}
\fourIdx{W}{-\infty}{\nu}{x}{\mathcal{D}}f(x)=
\fourIdx{RL}{-\infty}{\nu}{x}{\mathcal{D}}f(x)
=\fourIdx{C}{-\infty}{\nu}{x}{\mathcal{D}}f(x).
\end{equation}
One also has $\mathcal{W}^{\nu}e^{\lambda x}
=\lambda^{\nu} e^{\lambda x}$.
This property is convenient for the Fourier transform
$\mathcal{F}\big[f(x)\big](k)=\hat{f}(k)$,
which yields
\begin{equation}\label{A8-3}
\mathcal{F}\left[\mathcal{W}^{\nu}f(x)\right](k)=(-ik)^{\nu}\hat{f}(k).
\end{equation}

The fractional derivation with the
fixed lower limit is also called the left fractional derivative.
One can introduce the right fractional derivative, where
the upper limit $b$ is fixed and $b>x$. For example,
the formal definition of the right fractional derivative is
\begin{equation}\label{A8-4}
\fourIdx{}{x}{\nu}{b}{\mathcal{D}}f(x)
=\frac{1}{\Gamma(-\nu)}
\int_x^{b}\frac{f(y)dy}{(y-x)^{-(1+\nu)}},
\end{equation}
which is regularized  in either Riemann-Liouville \eqref{A3}
or Caputo \eqref{A4}  forms.

The Riesz fractional integral \cite{SaKiMa93} on the finite
interval $[a,b]$ is
\begin{equation}\label{A8-5}
\frac{1}{2\Gamma(\nu)\cos\tfrac{\nu\pi}{2}}
\int_a^b\frac{f(y)dz}{\lvert x-y\rvert^{1-\nu}}\, ,
\end{equation}
where $a\le x\le b$ and $0<\nu<1$. It can be represented
as the sum of the left and right Riemann-Liouville fractional
integrals
\begin{equation}\label{A8-6}
\int_a^x\frac{f(y)dy}{(x-y)^{1-\nu}}+
\int_x^b\frac{f(y)dy}{(y-x)^{1-\nu}}.
\end{equation}
Consequently, the Riesz fractional
derivative  $(-\Delta)^{\frac{\nu}{2}}$ on the entire
$x$-axis can also be represented by the Weyl derivatives
\begin{equation}\label{A8-7}
(-\Delta)^{\frac{\nu}{2}}\equiv \fourIdx{}{\infty}{\nu}{\lvert x\rvert}{\mathcal{D}}f(x)
=\frac{1}{2\cos\tfrac{\nu\pi}{2}}
\left[\fourIdx{W}{-\infty}{\nu}{x}{\mathcal{D}}f(x) +\fourIdx{W}{-\infty}{\nu}{-x}{\mathcal{D}}f(-x)\right].
\end{equation}

We also use a ``group property''
\begin{equation}\label{A10}
\clD_t^{-\nu}\left[\clD_t^{-\mu}f(t)\right]
=\clD_t^{-(\nu+\mu)}f(t)
=D_t^{-\mu}\left[\clD_t^{-\nu}f(t)\right],
\end{equation}
which is based on Dirichlet's formula
$\int_a^bdx\int_a^xf(x,y)dy=\int_a^bdy\int_y^bf(x,y)dx$ .
Another important property, used in the analysis, is
a combination of fractional integro-differentiation
\begin{equation}\label{A11}
\clD_t^{1-\beta}\, \partial^{\beta}f(t)=
D_tI_t^{\beta}I_t^{1-\beta}D_tf(t)
=D_t[f(t)-f(0)]=\frac{df(t)}{d\,t}\, ,
\end{equation}
where $0<\beta<1$. Using Eq. \eqref{A11}, one defines a fractional Taylor
expansion \cite{TrRiBo99} as follows
\begin{multline}\label{A12}
f(t+\vep)-f(t)=\fourIdx{}{t}{\alpha}{t+\vep}{I}
\fourIdx{C}{t}{\alpha}{t+\vep'}{\clD}f(t+\vep)= \\
=\frac{1}{\Gamma(\alpha)}\int_t^{t+\vep}(t+\vep-t')^{\alpha-1}
\fourIdx{C}{t}{\alpha}{t'}{\clD}f(t')dt'\, .
\end{multline}
Applying integration by parts $N$ times, we have from \eqref{A12}
\begin{multline}\label{A13}
f(t+\vep)=f(t)+\frac{\vep^{\alpha}}{\Gamma(\alpha+1)}\partial_t^{\alpha}f(t)
+\frac{1}{\Gamma(\alpha+1)}\int_t^{t+\vep}(t+\vep-t')^{\alpha}\,
\fourIdx{C}{t}{\alpha+1}{t'}{\clD}f(t')dt'=\\
\dotsb =f(t)+\sum_{n=0}^N\frac{\vep^{\alpha+n}}{\Gamma(\alpha+n+1)}
D^n_t\partial_t^{\alpha}f(t) + \mathrm{O}(\vep^{\alpha+n+1})\, .
\end{multline}

\subsection{Mittag-Leffler function and Fox $H$-function}
\label{sec:app_A1}

The fractional derivative of an exponential function can be
calculated by virtue of the Mittag-Leffler
function. Therefore, applications of
the Riemann-Liouville fractional derivative to the exponential
results in the Mittag-Leffler function
$\clD_t^{\alpha}e^{\lambda t}=t^{-\alpha}E_{1,1-\alpha}(\lambda t)$,
where the Mittag--Leffler function is (see {\em e.g.},
Refs. \cite{Po99,BaEr55})
\begin{subequations}\label{A9}
\begin{align}
E_{\nu,\beta}(z)=\frac{1}{2\pi i} \int_{\clC}\frac{\tau^{\nu-\beta}
e^{\tau}}{\tau^{\nu}-z}d\tau =
 \label{A9a}\\
=\sum_{k=0}^{\infty}\frac{z^k}{2\pi i} \int_{\clC}
\tau^{-k\nu-\beta}e^{\tau}
=\sum_{k=0}^{\infty}
\frac{z^k}{\Gamma(\nu k+\beta)} \label{A9b}\, .
\end{align}
\end{subequations}
Here $\nu,\beta>0$ and $\clC$ is a suitable contour
of integration, starting and
finishing at $-\infty$ and encompassing a circle $|\tau|\le
|z|^{\frac{1}{\nu}}$ in the positive direction.
For $\beta=1$ it also defines $E_{\nu,1}(z)\equiv E_{\nu}(z)$.

Another ubiquitous function in the theory of
fractional differential equations is the Fox $H$-function.
A detailed description of the Fox $H$-function and
its application can be found in Refs.~\cite{MSH2010,MaSa78}.
The Fox $H$-function is defined in terms of the Mellin-Barnes integral
\begin{multline}\label{A14}  
H_{p,q}^{m,n}(z)=H_{p,q}^{m,n}\left[z\left\lvert\begin{split}&(a_p,A_p)\\
&(b_q,B_q)\end{split}\right.\right]=\\
H_{p,q}^{m,n}\left[z\left\lvert\begin{split}&(a_1,A_1),\dotsc,(a_p,A_p)\\
&(b_1,B_1),\dotsc,(b_q,B_q)\end{split}\right.\right]=\frac{1}{2\pi i}
\int_{C}ds\,\theta(s)z^{-s}\, .
\end{multline}
Here
\begin{equation}\label{A15} 
\theta(s)=\frac{\prod_{j=1}^{m}\Gamma(b_j+B_js)
\prod_{j=1}^{n}\Gamma(1-a_j-A_js)}{
\prod_{j=m+1}^{q}\Gamma(1-b_j-B_js)\prod_{j=n+1}^{p}\Gamma(a_j+A_js)}\, ,
\end{equation}
where $n,m,p,q$ are non-negative integers with $0\leq n\leq p$,
$1\leq m\leq q$, $a_i,b_j \in \mathbb{C}$,
$A_i,B_j\in \mathbb{R}^{+}$, $i=1,\dotsc,p$, and $j=1,\dotsc,q$.
If the product is empty, then it is set equal to one.
The contour $C$ starts at $c- i\infty$, ends
at $c+ i\infty$, and separates the poles $\xi_{j,k}=-(b_j+k)/B_j$
of the gamma function $\Gamma(b_j+B_js)$
with  $j=1,\dotsc,m$ and $k=0,1,2,\dotsc$
from the poles $\chi_{i,k}=(1-a_i+k)/{A_i}$ of
the gamma function $\Gamma(1-a_i-A_is)$, $i=1,\dotsc,n$.

The Fox $H$-function relates to the exponential function and
the Mittag-Leffler function \cite{MSH2010,HMS2011,MaSa78}.
Evaluating the Mellin-Barnes integral as a sum of residues
in Eq. \eqref{A14}, we have
\begin{equation}\label{A19a}
H_{2,1}^{1,1}\left[-z\left\lvert\begin{split}&(0,1)\\
&(0,1),(1-\beta,\alpha)\end{split}\right.\right]
=\sum_{k=0}^{\infty}\frac{z^k}{\Gamma(k\alpha+\beta)}=
E_{\alpha,\beta}(z)\, ,   
\end{equation}
\begin{equation}\label{A19b}
H_{0,1}^{1,0}\left[z\left\lvert\begin{split}&{}\\
&(0,1)\end{split}\right.\right]=
\sum_{k=0}^{\infty}\frac{(-1)^k}{k!}z^k=e^{-z}\, .
\end{equation}
Here we used the Euler's reflection formula
$ \Gamma(p)\Gamma(1-p)=\frac{\pi}{\sin(p\pi)}$.

The Mellin transform is
$\mathcal{M}[f(t)](s)=\int_0^{\infty}f(t)t^{s-1}dt$,
and the contour integral in definition  (\ref{A14})
is the inverse Mellin transform.
Therefore, the Mellin transform of the Fox $H$-function yields
\begin{equation}\label{A22}   
\int_0^{\infty}dt\,t^{s-1}H_{p,q}^{m,n}
\left[ax\left\lvert\begin{split}
&(a_p,A_p)\\&(b_q,B_q)\end{split}\right.\right]
=a^{-s}\theta(s)\, ,
\end{equation}
where $\theta(s)$ is defined in Eq.\ (\ref{A14}).

The Mellin-cosine transform of the Fox H-function is
\begin{multline}\label{A22-24}
\int_0^{\infty}d\kappa\kappa^{\rho-1}\cos(\kappa x)
H_{p,q}^{m,n}\left[a\kappa^{\delta}
\left|\begin{array}{cc} (a_p\, ,A_p) \\ (b_q\, , B_q) \end{array}
\right.\right] \\
=\frac{\pi}{x^{\rho}}
H_{q+1,p+2}^{n+1,m}\left[\frac{x^{\delta}}{a}
\left|\begin{array}{cc} (1-b_q\, ,B_q),(\frac{1+\rho}{2}\, ,
\frac{\delta}{2}) \\
(\rho,\delta),(1-a_p\, ,A_p),(\frac{1+\rho}{2}\, ,\frac{\delta}{2})
\end{array}\right.\right] \, ,
\end{multline}
where
\begin{align*}
&{\rm Re}\left(\rho+\delta\cdot {\rm min}_{\substack{1\le j\le m}}(b_j/B_j)\right)
> 1 \, , \quad x^{\delta}>0\, \\
&{\rm Re}\left(\rho+\delta\cdot{\rm min}_{\substack{1\le j\le n}}((a_j-1)/A_j) \right) > 3/2\, , \quad  |{\rm arg}(a)| < \pi\tilde{\alpha}/2\, , \\
&\tilde{\alpha}= \sum^n_{j=1}A_j -\sum^p_{j=n+1}A_j +
\sum^m_{j=1}B_j - \sum^q_{j=m+1}B_j > 0\, .
\end{align*}
A combination of the Mellin and Laplace transforms also yields
a well known expression
\begin{equation}\label{A24}
\clM\left[\clL[f(t)](s)\right](1-p)=
\Gamma(1-p)\clM[f(t)](p)\, .
\end{equation}

\section{Quantization of fractional dynamics:
path integral approach}\label{sec:app_B}

Let us consider a Lagrangian function that contributes to
Green's function \eqref{ftse-5}.
The Lagrangian of a fractional ``free'' particle is
\begin{equation}\label{B-pi-1}
L(x,\partial^{\beta}_tx)=
\frac{1}{2}\left(\partial^{\beta}_tx\right)^2 \, .
\end{equation}
Now, let us introduce a fractional action by means of
fractional integration
\[S(T)=\frac{1}{\Gamma(\alpha)}
\int_{0}^TL(x,\partial^{\beta}_tx)(T-t)^{\beta-1}dt\, .\]
Following Feynman's heuristic arguments \cite{feynman},
the Green's function of the quantum evolution,
or the transition probability amplitude, for the wave
function can be constructed by
the path integral on the interval $T=t-t_0$.  
To perform this procedure, it
is instructive to introduce some random field $\xi(t)$,
and present the Green's function by means of the
complex Gaussian integral \cite{sebastian,CaSa08,wio}
\begin{equation}\label{B-pi-3}
\clG(x,T|x_0,0)=\int_{x_0,0}^{x_T,T}
\delta\left(\xi(t)-\partial^{\beta}_tx\right)
\exp\left[\frac{i}{2\hh\Gamma(\beta)}
\int_0^T[\xi(t)]^2(T-t)^{\beta-1}dt\right][d\xi(t)]\, ,
\end{equation}
which is an expectation value of the classical propagator.
The Dirac delta function ensures that the integration
accounts the trajectories, associated with the fractional Langevin equation
\begin{equation}\label{B-pi-4}
\partial^{\beta}_tx=\xi(t)\, .
\end{equation}

To proceed, we consider the fractional action in Eq. \eqref{B-pi-3}
as follows
\begin{equation}\label{B-pi-5}
S[\xi(t)]=\frac{1}{2\Gamma(\beta)}
\int_0^T[\xi(t)]^2(T-t)^{\beta-1}dt\, .
\end{equation}
Since the integral in Eq. \eqref{B-pi-3} is Gaussian,
the stationary phase method
yields the exact integration. Therefore,
we perform the variable change
$\xi(t)=\xi_{\rm e}(t)+\eta(t)$, where
$\xi_{\rm e}(t)$ is an extremum solution.
This corresponds to Eq. \eqref{B-pi-4}, which also
implies a constraint condition for $\eta(t)$. Namely, applying
the property \eqref{A11} to Eq. \eqref{B-pi-4}, we have
for the extremum path $\xi_{\rm ex}(t)$
\begin{equation}\label{B-pi-6}
x_t-x_0=\clD_t^{-\beta}\xi_{\rm e}(t)\equiv
\fourIdx{}{0}{\beta}{t}{I}\xi_{\rm e}(t)\, ,
\end{equation}
which also yields the constraint for the
deviation $\clD_t^{-\beta}\eta(t)=0$,
or $\delta\left(\clD_{t}^{-\beta}\eta(t)\right)$.
The path integral \eqref{B-pi-3} with respect to $\xi(t)$
reduces to the path integral with respect to $\eta(t)$.
The latter reads
\begin{multline}\label{B-pi-7}
\clG(x,T|x_0,0)=e^{\frac{i}{\hh}S_{\rm e}[\xi_{\rm e}]}
\int_{x_0,0}^{x_T,T}
\delta\left(D_{t}^{-\beta}\eta(t)\right)
e^{\frac{i}{\hh}S[\eta(t)]}[d\eta(t)] \equiv \\
\equiv F(T) e^{\frac{i}{\hh}S_{\rm e}[\xi_{\rm e}]} \, .
\end{multline}
The multiplier $F(T)\equiv \mathbb{A}^{-1}$ is the path integral
with respect to $\eta(t)$, which can be
obtained from the normalization condition
\begin{equation}\label{B-pi-8}
\int_{-\infty}^{\infty}\clG(x,T|x_0,0)dx_T=1\, , \quad \forall T\, .
\end{equation}
The extremum action $S_{\rm e}[\xi_{\rm e}]$  is found from a standard
technique of Lagrange multipliers. Following Ref. \cite{wio}, we add a
zero to the extremum action.  Taking into account Eqs.
\eqref{B-pi-4} and \eqref{A10}, we have
\begin{equation}\label{B-pi-9}
S[\xi_{\rm e}(t)\, ,\lambda]=\tfrac{1}{2\Gamma(\beta)}
\int_0^T\left[\xi^2_{\rm e}(t)+
2\lambda\left(\partial_t^{\beta}x-\xi_{\rm e}(t)\right)
\right](T-t)^{\beta-1}dt \, .
\end{equation}
Variation over $\lambda$ yields Eq. \eqref{B-pi-4},
and correspondingly Eq. \eqref{B-pi-6}, while variation over $\xi$ yields
\begin{equation}\label{B-pi-10}
\frac{1}{\Gamma(\beta)}\int_0^T\left(\xi_{\rm e}(t)- \lambda
\right)(T-t)^{\beta-1}\delta\xi(t) dt=0\, .
\end{equation}
Since $\delta\xi(t)$ is arbitrary, the solution to Eq. \eqref{B-pi-10}
for the extremum path is $\xi_{\rm e}(t)=\lambda$ and
from Eq. \eqref{B-pi-6} we obtain
\begin{equation}\label{B-pi-11}
 x_T-x_0=\fourIdx{}{0}{\beta}{T}{I}\lambda=
 \frac{\lambda T^{\beta}}{\Gamma(\beta+1)}\, ,
\end{equation}
which yields the Lagrange multiplier and the extremum path
\begin{equation}\label{B-pi-12}
\xi_{\rm e}(t)=\lambda=\Gamma(\beta+1)(x_T-x_0)/T^{\beta}\, .
\end{equation}
Correspondingly, the extremum action reads
\begin{equation}\label{B-pi-11-12}
S_{\rm e}[\xi_{\rm e}(t)]=\frac{1}{2\Gamma(\beta)}
\int_0^T[\xi_{\rm e}(t)]^2(T-t)^{\beta-1}dt \\
=\frac{m_{\beta}(x_T-x_0)^2}{2T^{\beta}\Gamma(\beta+1)}\, ,
\end{equation}
where $m_{\beta}=[\Gamma(\beta+1)]^2$ is an effective dimensionless mass
due to the fractional time dynamics. For $\beta=1$, it is $m_{\beta}=1$.
Inserting it in Eq. \eqref{B-pi-7},
and accounting the normalization condition \eqref{B-pi-8},
one obtains the Green's function as follows
\begin{subequations}\label{B-pi-13}
\begin{align}
\clG(x_T,T|x_0,0)= F(T)\exp\left[
\frac{im_{\beta}(x_T-x_0)^2}{2\hh\Gamma(\beta+1) T^{\beta}}
\right]\, , \label{B-pi-13a} \\
F(T)\equiv \mathbb{A}^{-1}=\frac{1}{\sqrt{2i\hh\Gamma(\beta+1)
\pi T^{\beta}/m_{\beta}}}\, . \label{B-pi-13b}
\end{align}
\end{subequations}


\begin{thebibliography}{0}

\bibitem{oldham} K. B. Oldham and J. Spanier,
\textit{The Fractional
Calculus} (Academic Press, Orlando, 1974).

\bibitem{Po99} I. Podlubny, \textit{Fractional Differential Equations}
(Academic Press, San Diego, 1999).

\bibitem{shlesinger} E. W. Montroll and M. F. Shlesinger,
in \textit{Studies in Statistical Mechanics},
eds. J. Lebowitz and E. W. Montroll
(Noth--Holland Physics Publishing, Amsterdam, 1984), Vol. 11, pp 1-122.

\bibitem{klafter} R. Metzler and J. Klafter,
\textit{Phys. Rep}. \textbf{339}, 1 (2000).

\bibitem{Moti1} R. Bekenstein, R. Schley, M. Mutzafi, C. Rotschild,
and M. Segev, \textit{Nature Phys.} \textbf{11}, 872 (2015).

\bibitem{Rivka1} R. Bekenstein, J. Nemirovsky, I. Kaminer,
and M. Segev, \textit{Phys. Rev. X} \textbf{4}, 011038 (2014).

\bibitem{Berry79} M. V. Berry, \textit{J. Phys. A: Math. Gen.}
\textbf{12}, 781 (1979). 

\bibitem{SeSoDu2012} M. Segev, M Solja\v{c}i\'{c}, and J. M. Dudley,
\textit{Nature Phot.} \textbf{6}, 209 (2012).

\bibitem{Moti2} T. Schwartz, G. Bartal, S. Fishman, and M. Segev,
\textit{Nature} {\bf 446}, 52 (2007).

\bibitem{LeKrFiSe2012} L. Levi, Y. Krivolapov, S. Fishman,
and M. Segev, \textit{Nature Phys.} \textbf{8}, 912 (2012).

\bibitem{iom2015} A. Iomin, \textit{Phys. Rev. E},
\textbf{92}, 022139 (2015).

\bibitem{laskin1} N. Laskin, 
\textit{Chaos} \textbf{10}, 780 (2000).

\bibitem{west} B. J. West, 
\textit{J. Phys. Chem. B} \textbf{104}, 3830 (2000).

\bibitem{feynman} R. P. Feynman and A. R. Hibbs,
\textit{Quantum mechanics and path integrals}
(McGraw-Hill, New York, 1965).

\bibitem{kac} M. Kac, \textit{Probability and related topics in physical
sciences} (Interscience, New York, 1959).

\bibitem{ZT} V. E. Tarasov and G. M. Zaslavsky,
\textit{Comm. Nonlinear Scien. Num. Simul.} \textbf{13}, 248, (2008).

\bibitem{Longhi} S. Longhi,
\textit{Optics Lett.} \textbf{40}, 1117 (2015).

\bibitem{Laskin2} N. Laskin, \textit{Phys. Lett. A} \textbf{268},
298 (2000); \textit{Phys. Rev. E} \textbf{66}, 056108 (2002).

\bibitem{relat_oscill} K. Kowalski and J. Rembielinski,
\textit{Phys. Rev. A} \textbf{81}, 012118 (2010);
J. L\"orinczi and J. Malecki, \textit{J. Differential Equations}
\textbf{253}, 2846 (2012).

\bibitem{GVega} J. C. Guti\'errez-Vega,
\textit{Opt. Lett.} \textbf{32}, 1521 (2007);
\textit{Opt. Express} \textbf{15}, 6300 (2007).

\bibitem{Y_Zhang} Y. Zhang,
\textit{Phys. Rev. Lett.} \textbf{115}, 180403 (2015).

\bibitem{L_Zhang2} L. Zhang \textit{et al.},
\textit{Opt. Express} \textbf{24}, 14406 (2016).

\bibitem{L_Zhang1} L. Zhang \textit{et al.},
\textit{ Opt. Express} \textbf{27}, 27936 (2019).

\bibitem{HuMcD2005} J. G. Huang and G. S. McDonald,
\textit{Phys. Rev. Lett.} \textbf{94}, 174101 (2005).

\bibitem{Sha2002} V. M. Shalaev,
\textit{Nonlinear Optics of Random Media:
Fractal Composites and Metal-Dielectric Films}
(Springer, New York, 2000).

\bibitem{jad2014} A. Jadczyk,
\textit{Quantum fractals: from Heisenberg's uncertainty to Barnsley's fractality} (WS, Singapore, 2014).

\bibitem{SiSu2019} R. T. Sibatov and H. G. Sun,
\textit{Fractal Fract.} \textbf{3}, 47 (2019).

\bibitem{wo2010} J.-N. Wu, C.-H. Huang, S.-C. Cheng, and W.-F. Hsieh,
\textit{Phys. Rev. A} \textbf{81}, 023827 (2010).

\bibitem{iom2009} A. Iomin,
\textit{Phys. Rev. E} \textbf{80}, 022103 (2009).

\bibitem{IS16} A. Iomin and T. Sandev,
\textit{Math. Model. Nat. Phenom} \textbf{11}, 51 (2016).  

\bibitem{Bender} C. M. Bender, \textit{Rep. Prog. Phys.}
\textbf{70}, 947 (2007).

\bibitem{OPTICS} C. E. R\"uter, \textit{et al.},
\textit{Nat. Phys.} \textbf{6}, 192 (2010);
S. Longhi, \textit{Phys. Rev. A} \textbf{82}, 031801 (2010);
A. Regensburger, \textit{et al.}, \textit{Nature}
\textbf{488}, 167 (2012).

\bibitem{Ta08} V. E. Tarasov,
\textit{Quantum Mechanics of Non-Hamiltonian and Dissipative Systems} (Elsevier, Amsterdam, 2008).

\bibitem{La2017} N. Laskin,
\textit{Chaos, Solitons} \& \textit{Fractals} \textbf{102},
16 (2017).  

\bibitem{iom2019} A. Iomin,
\textit{Chaos, Solitons} \& \textit{Fractals X} \textbf{1},  100001, (2019).

\bibitem{naber} M. Naber,
\textit{J. Math. Phys.} \textbf{45}, 3339 (2004).  

\bibitem{list1} S. Wang, M. Xu, \textit{J. Math. Phys.}
\textbf{48}, 043502 (2007).

\bibitem{list2} J. Dong, M. Xu,
\textit{J. Math. Anal. Appl.} \textbf{344}, 1005 (2008). 

\bibitem{iom2011} A. Iomin,  {\it Chaos, Solitons} \& {\it Fractals}
\textbf{44}, 348 (2011). 

\bibitem{list3} S. Bayin,
\textit{J. Math. Phys.} \textbf{54}, 012103 (2013).

\bibitem{list4} P. G{\'o}rka, H. Prado, and J. Trujillo,
\textit{Integr. Equ. Oper. Theory}  \textbf{87}, 1 (2017), 

\bibitem{AcYaHa13} B. N. Narahari Achar, B. T. Yale, and J. W. Hanneken,
\textit{Adv. Math. Phys.} \textbf{2013}, 290216 (2013).

\bibitem{SaPeLe14} T. Sandev, I. Petreska, and E. K. Lenzi,
\textit{J. Math. Phys.} \textbf{55}, 092105 (2014).

\bibitem{stone} M. H. Stone,
\textit{Ann. Math.} \textbf{33}, 643 (1932).  

\bibitem{Fibich} G. Fibich, \textit{The Nonlinear Schr\"odinger
Equation: Singular Solutions and Optical Collapse}
(Springer, Cham, 2015).

\bibitem{leontovich} M. A. Leontovich, Izv. USSR Ac.Sc.,
Phys. \textbf{8}, 16, 1944 (in Russian).

\bibitem{khohlov} R. V. Khohlov,  Radiotech. and Elrctron.
\textbf{6}, 1116, 1961 (in Russian).

\bibitem{kliatskin} V. I. Klyatskin,
\textit{Stochastic Equations: Theory and Applications in
Acoustics, Hydrodynamics, Magnetohydrodynamics,
and Radiophysics} (Springer, Heidelberg, 2015).

\bibitem{tappert} E. D. Tappert, in
\textit{Lectures Notes in Physics} \textbf{70},
eds. J. B. Keller and J. S. Papadakis
(Springer, New York, 1977) p. 224.

\bibitem{LevyLens} P. Barthelemy, J. Bertolotti, D. S. Wiersma,
\textit{Nature} \textbf{453}, 495 (2008).

\bibitem{nirD} Y. Sagi, M. Brook, I. Almog, N. Davidson,
\textit{Phys. Rev. Lett.} \textbf{108}, 093002 (2012).

\bibitem{zoller} S. Marksteiner, K. Ellinger, P. Zoller,
\textit{Phys. Rev. A} \textbf{53}, 3409 (1996).

\bibitem{eli1} D. A. Kessler, E. Barkai,
\textit{Phys. Rev. Lett.} \textbf{108}, 230602 (2012).

\bibitem{eli2} A. Dechant, E. Lutz, D. A. Kessler, E. Barkai,
\textit{Phys. Rev. Lett.} \textbf{107}, 240603 (2011).

\bibitem{iom2012} A. Iomin,
\textit{Phys. Rev. E} \textbf{86}, 032101 (2012).

\bibitem{BeBa79} M. V. Berry and N. L. Balazs,
\textit{Am. J. Phys.} \textbf{47}, 264 (1979).

\bibitem{MSH2010}
A. M. Mathai, R. K. Saxena, H. J. Haubold.
{\it The H-function: Theory and Applications} (Springer, New York, 2010).

\bibitem{LaZa2006} N. Laskin, and G. M. Zaslavsky,
\textit{Physica A} \textbf{368}, 38 (2006).  

\bibitem{Ta2015} V. E. Tarasov, \textit{Nonlinear Dyn.}
\textbf{80}, 1665 (2015).  

\bibitem{TaZa2005} V. E. Tarasov and G. M. Zaslavsky,
\textit{Physica A} \textbf{354}, 249 (2005). 

\bibitem{MiRa2005} A. V. Milovanov and J. J. Rasmussen,
\textit{Phys. Lett. A} \textbf{337}, 75 (2005).  

\bibitem{TaZa2006a} V. E. Tarasov and Zaslavsky,
\textit{Chaos} \textbf{16}, 023110 (2006).

\bibitem{TaZa2006b}  V. E. Tarasov and Zaslavsky,
\textit{Commun. Nonlinear Sci. Numer. Simul.} \textbf{11},
885 (2006).  

\bibitem{ZaEdTa2007} G. M. Zaslavsky, M. Edelman, and V. E. Tarasov,
\textit{Chaos} \textbf{17}, 043124 (2007).

\bibitem{CaSi2007} L. Caffarelli and L. Silvestre,
\textit{Comm. Partial Diff. Equat.} \textbf{32}, 1245 (2007). 

\bibitem{PaKoetal2020}
J. L. Padgett, E. G. Kostadinova, C. D. Liaw, K. Busse, L. S. Matthews,
and T. W. Hyde, \textit{J. Phys. A: Math. and Theor.}
\textbf{53}, 135205 (2020).

\bibitem{CaTa2010} X. Cabr\'e and J. Tan,
\textit{Advances in Math.} \textbf{224}, 2052 (2010). 

\bibitem{Ta2011} J. Tan, \textit{Calc. Var.}
\textbf{42}, 21 (2011).   

\bibitem {Anderson} P. W. Anderson, \textit{Phys. Rev.} {\bf 109},
1492 (1958).

\bibitem{LGP} I. M. Lifshits, S. A. Gredeskul, and L. A.
Pastur, {\it Introduction to the theory of disordered systems}
(Wiley-Interscience, New York, 1988).

\bibitem{lahini} Y. Lahini {\it et al}.,
\textit{Phys. Rev. Lett.} {\bf 100}, 013906 (2008).

\bibitem{SanchAspect} L. Sanchez-Palencia, {\it et al}.,
\textit{Phys. Rev. Lett.} {\bf  98}, 210401 (2007).

\bibitem{aspect} J. Billy, {\it et al}.,
\textit{Nature} {\bf 453}, 891 (2008).

\bibitem{fks}
S. Flach, D. O. Krimer, and Ch. Skokos.
\textit{Phys. Rev. Lett.} \textbf{102}, 024101 (2009).

\bibitem{iom2010} A. Iomin.
\textit{Phys. Rev. E} \textbf{81}, 017601 (2010).

\bibitem{MiIo2012} A. V. Milovanov, A. Iomin.
\textit{Europhys. Lett.} \textbf{100}, 10006 (2012).

\bibitem{basko} D. M. Basko, \textit{Ann. Phys.} \textbf{326}, 1577 (2011).

\bibitem{SeMaSk2018} B. Senyange, B. Many Manda, and Ch. Skokos,
\textit{Phys. Rev. E} \textbf{98}, 052229 (2018).

\bibitem{iom2016} A. Iomin, \textit{Chaos, Solitons} \& {\it Fractals}
\textbf{93}, 64 (2016). 

\bibitem{schiff} L. I. Schiff, \textit{Quantum Mechanics}
(McGraw-Hill, New York, 1968).

\bibitem{MiIo2017} A. V. Milovanov and A. Iomin,
\textit{Phys. Rev. E} \textbf{95}, 042142 (2017).

\bibitem{Aksel_etal2014} G. M. Akselrod, et al.,
\textit{Nature Commun.} \textbf{5}, 3646 (2014).

\bibitem{Penrose1} R. Penrose,
\textit{Phil. Trans. R. Soc.} \textbf{356}, 1927 (1998).

\bibitem{Diosi} L. Di\'osi, \textit{Phys. Lett. A} \textbf{105}, 199 (1984).

\bibitem{Penrose2} R. Penrose, \textit{Gen. Relativ. Gravit.}
\textbf{28}, 581 (1996).

\bibitem{Penrose3} R. Penrose, \textit{Found. Phys.}
\textbf{44}, 557 (2014).

\bibitem{Bahrami2014} M. Bahrami, A. Grossardt, S. Donadi,
and A. Bassi,
\textit{New J. Phys.} \textbf{16}, 115007 (2014).

\bibitem{guisti2020} A. Giusti, R. Garrappa, and G. Vachon,
\textit{Eur. Phys. J. Plus} \textbf{135}, 798 (2020).

\bibitem{Preprint}  G. P. Berman, A. M. Iomin, A. R. Kolovsky,
and N. N. Tarkhanov,
Report No. 377F, Kirensky Institute of Physics, Krasnoyarsk,
1986 (in Russian, unpublished).

\bibitem{BeKo84} G. P. Berman and A. R. Kolovskii,
\textit{Sov. Phys. JETP} \textbf{50}, 1116 (1984).

\bibitem{ErDeSiBu10} H. Ertik, D. Demirhan, H. Sirin, and
F. B$\ddot{u}$y$\ddot{u}$kkili\c{c},
\textit{J. Math. Phys.} \textbf{51}, 082102 (2010).

\bibitem{SiTs82} Ya. A. Sinitsyn and V. M. Tsukernik,
\textit{Phys. Lett. A} \textbf{90}, 39 (1982). 

\bibitem{BeIo85} G. P. Berman and A. M. Iomin,
\textit{Sov. Phys. JETP} \textbf{62}, 544 (1985). 


\bibitem{iom17} A. Iomin,
\textit{Computers} \& {\it Mathematics with Applications} \textbf{73},
914 (2017). 

\bibitem{zacher} R. Zacher, \textit{Math. Ann.} \textbf{356}, 99 (2013).

\bibitem{Calcagni} G. Calcagni, 
\textit{Phys. Rev. Lett.} \textbf{104}, 251301 (2010).

\bibitem{ElNab05} R. A. El-Nabulsi,
\textit{Fiz. A} \textbf{14}, 289 (2005).  

\bibitem{schulman} L. Schulman, \textit{Techniques and applications of
path integration}  (Wiley, New York, 1981).



\bibitem{SaKiMa93} S. G. Samko, A. A. Kilbas,  and O. I. Marichev,
\textit{Fractional integrals and derivatives: theory and applications} (Gordon and Breach, New York, 1993).

\bibitem{TrRiBo99} J. Truilljo, M. Rivero, and B. Bonilla,
\textit{J. Math. Anal.} \textbf{231}, 255 (1999).  

\bibitem{BaEr55} H. Bateman and A. Erd\'elyi,
\textit{Higher transcendental functions}, Vol. 3
(McGraw-Hill, New York, 1955).

\bibitem{HMS2011} H. J. Haubold A. M. Mathai, R. K. Saxena,
\textit{J. of Appl. Math.} \textbf{2011}, 298628 (2011).

\bibitem{MaSa78} A. M. Mathai and R. K. Saxena, {\it The H-Function with Applications in Statistics and Other Disciplines}
(John Wiley \& Sons, New York,  1978).

\bibitem{sebastian} K. L. Sebastian,
\textit{J. Phys. A: Math. Gen.} \textbf{28}, 435 (1995). 

\bibitem{CaSa08} I. Calvo and R. S\'anchez,
\textit{J. Phys. A: Math. Theor.} \textbf{41}, 282002 (2008).

\bibitem{wio} H. Wio, \textit{Path Integrals For Stochastic Processes}
(World Scientific, New Jersey, 2013).



\end{thebibliography}
\end{document}